\documentclass[a4paper, 11pt]{article}

\usepackage[utf8]{inputenc}
\usepackage[english]{babel}
\usepackage{amsthm, amsmath, amssymb}
\usepackage[T1]{fontenc}
\usepackage{fullpage}
\usepackage{lipsum}
\usepackage{authblk}
\usepackage{url} 
\usepackage{hyperref}
\usepackage{graphicx}
\usepackage{xcolor}
\usepackage{mathtools}
\usepackage{bbm}
\usepackage{tikz-cd,tikz-3dplot}
\usepackage{enumitem}
\usepackage{soul}
\usepackage{mathrsfs}
\usepackage{subcaption}
\usepackage{float}
\usepackage{longtable}
\usepackage{array}
\usepackage{algorithm}
\usepackage{algpseudocode}

\usepackage[maxbibnames=99, style=numeric]{biblatex} 

\newtheorem{assumption}{Assumption}[section]
\newtheorem{definition}[assumption]{Definition}
\newtheorem{theorem}[assumption]{Theorem}
\newtheorem{proposition}[assumption]{Proposition}

\newtheorem{lemma}[assumption]{Lemma}

\theoremstyle{remark}
\newtheorem{example}[assumption]{Example}
\theoremstyle{remark}
\newtheorem{remark}[assumption]{Remark}

\newcommand{\Fcal}{\mathcal{F}}
\newcommand{\Gcal}{\mathcal{G}}

\newcommand{\Ical}{\mathcal{I}}

\newcommand{\Lcal}{\mathcal{L}}
\newcommand{\Mcal}{\mathcal{M}}

\newcommand{\Scal}{\mathcal{S}}
\newcommand{\Tcal}{\mathcal{T}}

\newcommand{\Wcal}{\mathcal{W}}

\newcommand{\Prob}{\mathscr{P}}
\newcommand{\Mart}{\mathscr{M}}

\newcommand{\EE}{\mathbb{E}}
\newcommand{\FF}{\mathbb{F}}

\newcommand{\NN}{\mathbb{N}}

\newcommand{\PP}{\mathbb{P}}

\newcommand{\RR}{\mathbb{R}}

\newcommand{\supp}{{\textup{supp}}}
\newcommand{\co}{{\textup{co}}}

\newcommand{\Law}{{\textup{Law}}}
\newcommand{\interior}{{\textup{int}}}
\newcommand{\MCov}{{\textup{MCov}}}

\newcommand{\RRd}{{\mathbb R^d}}
\newcommand{\RRdd}{{\mathbb R^{d\times d}}}
\newcommand{\aff}{{\operatorname{aff}}}
\newcommand{\ri}{{\operatorname{ri}}}

\makeatletter
\renewcommand{\p@enumii}{}            

\makeatother

\newcounter{stepimg}

\newcommand{\stepfig}[2]{%
  \refstepcounter{figure}\label{#1}\setcounter{stepimg}{0}%
  \begin{longtable}{|@{\hspace{.5em}}
  >{\centering\arraybackslash}p{.32\textwidth}@{\hfill}
  >{\centering\arraybackslash}p{.32\textwidth}@{\hfill}
  >{\centering\arraybackslash}p{.32\textwidth}
  @{\hspace{1em}}|}
  \hline
  \multicolumn{3}{|@{\hspace{.5em}}l@{\hspace{.5em}}|}{%
    \rule{0pt}{1.5em}\textbf{Figure~\thefigure: #2}%
  }\\[0.5em]
  \hline
  \noalign{\vskip .25em}
}

\newcommand{\closestepfig}{%
  \hline
  \end{longtable}%
}

\title{Mimicking Volatilities under Marginal Constraints via $q_x$-Bass Martingales: A Parallel Computing Approach}
\author{Antonio Marini}

\affil{Department of Mathematics,\\ ETH Z\"{u}rich, Switzerland \\ \texttt{antonio.marini@math.ethz.ch}}

\begin{document}
\maketitle

\begin{abstract}
We study the calibration of continuous-time martingale models to observed derivative prices via a dynamic martingale optimal transport problem that penalizes deviations from the volatility of a prescribed reference model. Using $q_x$-Bass martingales, we approximate this problem through a sequence of discrete-time problems and prove that the continuous-time interpolations of their optimizers are asymptotically optimal. Building on the theory of $q_x$-Bass martingales, we develop numerical methods for the discrete-time problems that do not require full marginal distributions at calibration maturities.
\end{abstract}

\section{Introduction}

Constructing a pricing model consistent with observed market prices of financial derivatives is a central problem in mathematical finance. Classical approaches usually require selecting a class of stochastic models and then calibrating the associated parameters or coefficients to market data. Prominent examples include parametric stochastic volatility models, such as the Heston model \cite{Heston1993}, local volatility models, such as Dupire's local volatility model \cite{Du94}, and local-stochastic volatility models \cite{He09,GuHe11}, which combine the calibration flexibility of local volatility with the richer dynamics of stochastic volatility. While these approaches are widely used, they impose specific structural assumptions on the underlying price dynamics.

An alternative approach is provided by semimartingale optimal transport, where one selects, among all models consistent with market prices, the one closest to a prescribed reference model according to a suitable cost functional. Developed by Guo--Loeper--Wang for the calibration of local-stochastic volatility models \cite{GuLoWa22}, this methodology has been extended to joint SPX/VIX calibration \cite{GuLoObWa22} and stochastic interest-rate models \cite{JoLoOb26}. These approaches rely on PDE formulations and duality arguments involving Hamilton--Jacobi--Bellman equations.

When discounted asset prices are modeled as martingales, the same selection principle can be formulated within martingale optimal transport.  In this direction, Guo--Loeper--Wang \cite{GuLoWa19} proposed an optimal transport approach to local volatility calibration.

Another important development is represented by the martingale Benamou--Brenier formula \cite{BaBeHuKa20}, which selects, among Brownian martingales with prescribed initial and terminal marginals, the one whose dynamics are closest to Brownian motion. Its optimizers, originally called stretched Brownian motions, are now commonly referred to in the literature as Bass martingales, since they are a generalization of the classical construction introduced by Bass \cite{Ba83} to solve the Skorokhod embedding problem. Building on this construction, Conze and Henry-Labord\`ere \cite{CoHe21} introduced the Bass local volatility model, which allows for exact calibration to risk-neutral marginal distributions at finitely many maturities.
Unlike Dupire's approach, it does not require interpolation of option prices across maturities. Moreover, its calibration can be implemented through a simple fixed-point iteration, without solving PDEs, for which rigorous convergence results have been proved \cite{AcMaPa23, HaJoLoObPa26}.

The theory of Bass martingales has since been studied extensively, both for its structural and duality properties \cite{BaScTs23,Ts24} and for its numerical aspects and applications \cite{CoHe21,AcMaPa23,HaJoLoObPa26,AcGaMaPa26, BeHaPa26}. Furthermore, several works have extended the construction beyond the Brownian reference. The $q$-Bass martingales introduced by Tschiderer \cite{Ts24} replace the Gaussian reference distribution with a general probability measure $q$. Subsequent work by Acciaio and Marini \cite{AcMa26_semidiscrete,AcMa26_general} established existence and convergence results in dimension one and extended the martingale Benamou--Brenier formula to additive reference processes, including L\'evy processes. 
A further generalization is given by $q_x$-Bass martingales \cite{Ma26_qxBass}, where the reference measure is replaced by a family of probability kernels $(q_x)_{x\in\RRd}$. These martingales, whenever they exist, are optimizers of the weak martingale optimal transport problem
\begin{equation}\label{eq:qx_WOT}
    WT_{\Ical}^{q_\cdot}(\rho, \zeta) := \inf_{\pi\in\Mart(\rho,\zeta)} \int_{\RRd}\Wcal_2^2(\pi_x, q_x) \, \rho(dx),
\end{equation}
where $\Mart(\rho,\zeta)$ denotes the set of martingale couplings $\pi$ with marginals $\rho$ and $\zeta$, and $\pi_x$ is the conditional law of the second coordinate given the first.
The problem thus selects, among martingale couplings with prescribed marginals, those whose conditional distributions are, on average, closest to the reference kernels in quadratic Wasserstein distance. A key advantage of both $q$- and $q_x$-Bass martingales is that they preserve the algorithmic properties of the original Bass local volatility model \cite{Ma26_qxBass, AcMaPa23, HaJoLoObPa26, AcMa26_general}.

In the present paper, we use $q_x$-Bass martingales to construct continuous-time processes that approximately solve dynamic martingale optimal transport problems of the form
\begin{equation}\label{eq:mainPb}
\inf_{M\in\Mart_{[t_0,t_1]}} \mathbb{E}\left[\int_{t_0}^{t_1} \|\sigma_t - \widehat{\sigma}(t, M_t)\|_{\mathrm{HS}}^2\,dt\right].
\end{equation}
Here, $\widehat{\sigma}(t,x)$ denotes the diffusion coefficient of an $\RRd$-valued reference martingale, while $\Mart_{[t_0,t_1]}$ denotes the class of admissible Brownian martingales $(M_t)_{t\in[t_0,t_1]}$ that satisfy the prescribed calibration constraints and admit a representation $dM_t=\sigma_t\,dB_t$, with $(\sigma_t)_{t \in [t_0, t_1]}$ predictable and $(B_t)_{t \in [t_0, t_1]}$ a Brownian motion fixed throughout the paper and also driving the reference martingale. In particular, the candidate coefficient $\sigma_t$ need not be a function of $(t,M_t)$.
From a practical perspective, the main motivation behind \eqref{eq:mainPb} is to construct calibrated martingale models that mimic the volatility of a prescribed reference model by penalizing deviations between the corresponding diffusion coefficients.
Notice that the reference coefficient $\widehat{\sigma}$ is evaluated along the candidate process $M$, so that the discrepancy between the two instantaneous diffusion coefficients is measured under the distribution induced by the model being selected. In this way, the optimization favors models whose volatility remains close to the reference volatility in the regions of the state space they are most likely to visit.

To approximate \eqref{eq:mainPb}, we consider a time partition $\Tcal=\{t_0=s_1<\dots<s_{n+1}=t_1\}$ containing all maturities at which calibration constraints are imposed. We denote by $\Mcal_\Tcal$ the set of admissible chains of marginal distributions $(\eta_i)_{i=1}^{n+1}$, satisfying the prescribed constraints and increasing in convex order. Since these marginals do not uniquely determine a martingale model, we use \eqref{eq:qx_WOT} to select the transitions between consecutive maturities. More precisely, letting $q_x^{(i)}$ denote the transition kernel of the reference martingale over $[s_i,s_{i+1}]$, we consider the multimarginal weak optimal transport
\begin{equation}\label{eq:WOT}
MWT\big(\Mcal_\Tcal, (q_\cdot^{(i)})_{i=1}^n\big) := \inf_{(\eta_i)_{i=1}^{n+1}\in\Mcal_\Tcal}\sum_{i=1}^n WT_{\Ical}^{q_\cdot^{(i)}}(\eta_i, \eta_{i+1}).
\end{equation}

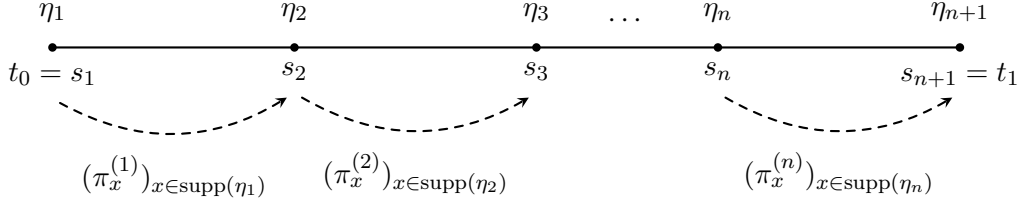
\begin{figure}[H]
\centering
\begin{tikzpicture}[>=stealth, line cap=round, line join=round]
    \coordinate (s1)   at (0,0);
    \coordinate (s2)   at (3.2,0);
    \coordinate (s3)   at (6.4,0);
    \coordinate (sn)   at (8.8,0);
    \coordinate (snp1) at (12.0,0);

    \draw[thick] (s1) -- (snp1);

    \foreach \p in {s1,s2,s3,sn,snp1}
        \fill (\p) circle (1.6pt);

    \node[above=6pt] at (s1) {$\eta_1$};
    \node[above=6pt] at (s2) {$\eta_2$};
    \node[above=6pt] at (s3) {$\eta_3$};
    \node[above=6pt] at (sn) {$\eta_n$};
    \node[above=6pt] at (snp1) {$\eta_{n+1}$};
    \node[below=2pt] (T1) at (s1) {$t_0=s_1$};
    \node[below=2pt] (T2) at (s2) {$s_2$};
    \node[below=2pt] (T3) at (s3) {$s_3$};
    \node[below=2pt] (Tn) at (sn) {$s_n$};
    \node[below=2pt] (Tnp1) at (snp1) {$s_{n+1}=t_1$};

    \node at (7.6,0.35) {$\cdots$};

    \draw[->, thick, dashed]
        ([xshift=1mm,yshift=-1mm]T1.south)
        to[bend right=30]
        node[midway, below=4pt]
        {$(\pi_x^{(1)})_{x\in\supp(\eta_1)}$}
        ([xshift=-1mm,yshift=-1mm]T2.south);
    
    \draw[->, thick, dashed]
        ([xshift=1mm,yshift=-1mm]T2.south)
        to[bend right=30]
        node[midway, below=4pt]
        {$(\pi_x^{(2)})_{x\in\supp(\eta_2)}$}
        ([xshift=-1mm,yshift=-1mm]T3.south);
        
    \coordinate (lastEndBase) at (Tn.south -| snp1);
    \draw[->, thick, dashed]
        ([xshift=1mm,yshift=-1mm]Tn.south)
        to[bend right=30]
        node[midway, below=4pt]
        {$(\pi_x^{(n)})_{x\in\supp(\eta_n)}$}
        ([xshift=-1mm,yshift=-1mm]lastEndBase);
\end{tikzpicture}

\caption{The optimization \eqref{eq:WOT} jointly selects the marginals $\eta_i$ and the martingale transition kernels $(\pi_x^{(i)})_{x\in\supp(\eta_i)}$, represented by the dashed arcs below, so that the calibration constraints defining $\Mcal_\Tcal$ are satisfied while minimizing the deviation from the reference kernels $q_\cdot^{(i)}$.}
\label{fig:WOT_scheme}

\end{figure}

This problem jointly selects the intermediate marginals $(\eta_i)_{i=1}^{n+1}$ and, through the corresponding $q_x$-Bass optimizers, the martingale transitions $((\pi_x^{(i)})_{x\in\supp(\eta_i)})_{i=1}^{n}$ between them (see Figure \ref{fig:WOT_scheme}).

The theoretical basis for this construction is recalled in Section~\ref{sec:qx-Bass}, where we also introduce the notion of continuous-time $q_x^{s,t}$-Bass extension. This extension preserves the optimal martingale coupling between the endpoints and provides the building block for constructing continuous-time processes from discrete $q_x$-Bass martingales. In Section~\ref{sec:dynamic-approximation}, we first establish the main properties of \eqref{eq:WOT}, including convexity and uniqueness of the optimal transitions between prescribed consecutive marginals (Proposition~\ref{prop:approximate-problem}). We then introduce the continuous-time Bass interpolation in Definition~\ref{def:continuous-time-q-Bass-interpolation}, obtained by gluing the corresponding $q_x^{s,t}$-Bass extensions along the partition. Our main approximation result, Theorem~\ref{thm:approximate-dynamic-MOT}, shows that, under suitable assumptions, the continuous-time martingales constructed from optimizers of \eqref{eq:WOT} approximate the optimal value of \eqref{eq:mainPb} as the mesh of the partition tends to zero, with a quantitative convergence estimate.

Furthermore, this approach offers two main advantages for model calibration. First, it allows for a broader class of reference models than those considered by the martingale Benamou--Brenier formula and its continuous-time generalizations to additive processes, while preserving their computational advantages \cite{AcMaPa23, HaJoLoObPa26, AcMa26_general, Ma26_qxBass}. This flexibility may help reduce or even remove discontinuities in the volatility of the selected model at calibration maturities, a limitation of the Bass local volatility model arising from the restrictive choice of Brownian motion as the reference process. Second, calibration constraints can be imposed directly in terms of observed derivative prices, without requiring the reconstruction of entire risk-neutral marginal distributions through the Breeden--Litzenberger formula \cite{BrLi78}. In particular, this avoids the need to interpolate option prices across strikes only to recover these distributions. The resulting framework thus allows for more general market constraints while offering an alternative to PDE-based calibration methods, with PDEs potentially needed only to compute the transition kernels of the reference process.

In Section~\ref{sec:numerical-algorithm}, we introduce numerical methods for the solution of \eqref{eq:WOT}. We first define a finite-dimensional discretization of the admissible marginals and establish its consistency with the original problem in Proposition~\ref{prop:finite-support-approximation}. Building on the strong duality theory of $q_x$-Bass martingales and the Generalized Martingale Sinkhorn Algorithm recalled in Section~\ref{sec:qx-Bass}, we derive subgradients of the objective from the optimal dual potentials (Proposition~\ref{prop:subgradient-F}). This allows us to formulate a preliminary approach based on standard methods for constrained convex optimization. We then focus on a projected subgradient method (Algorithm~\ref{alg:projected-subgradient}), whose convergence is established in Theorem~\ref{thm:convergence-projected-subgradient} via standard arguments. Unlike numerical approaches based on discretizing martingale couplings, such as that of Guo and Obłój~\cite{GuOb19}, our approach optimizes over the weights of the marginal distributions without explicitly treating the couplings between consecutive marginals as optimization variables. The additive structure of the objective allows the one-step transport problems, and hence the corresponding dual potentials, to be computed in parallel.

Finally, Section~\ref{sec:extensions} discusses further extensions and applications of our approach. We first introduce a relaxed formulation incorporating calibration errors through convex penalty terms, allowing for noisy or inconsistent market prices. We also consider an alternative formulation based on maximizing covariance with the reference model, highlighting its differences from the criterion used in \eqref{eq:mainPb}. This also provides applications to the Martingale Benamou–Brenier formula and its generalizations, particularly when only partial marginal information is available. We then discuss the possibility of enriching the reference dynamics with additional latent factors, as in stochastic volatility models, while imposing calibration constraints only on the asset-price component. A related construction also appears in the stochastic Knothe--Rosenblatt (SKR) model of Beiglböck, Hasenbichler, and Pammer \cite{BeHaPa26}. It uses stochastic volatility models as reference processes and is based on a generalization of the Bass construction named $R$-Bass martingale. In the SKR model, the price marginals are prescribed and the reference dynamics of the latent factor are preserved, while this formulation allows for partial marginal constraints and the dynamics of both components are optimized jointly.

\paragraph{Notation.}
Let $\Prob(\RRd)$ denote the space of probability measures on $\RRd$. For $p\in[1,\infty)$, we write $\Prob_p(\RRd)$ for the set of probability measures with finite $p$-th moment. We denote by $\lambda^d$ the Lebesgue measure on $\RRd$. For $\xi\in\Prob(\RRd)$, $\supp(\xi)$ denotes its support, and $\xi\ll\lambda^d$ means that $\xi$ is absolutely continuous with respect to $\lambda^d$.

For $\mu,\nu \in \Prob(\RRd)$, we denote by $\Pi(\mu,\nu)$ the set of probability measures on $\RRd\times\RRd$ whose first and second marginals are $\mu$ and $\nu$, respectively. Its elements are called couplings of $\mu$ and $\nu$. For $p \in [1,\infty)$ and $\xi,\zeta \in \Prob(\RRd)$, their $p$-Wasserstein distance is defined by
\[
    \Wcal_p(\xi, \zeta) := \inf_{\pi\in\Pi(\xi,\zeta)}\left(\int_{\RRd \times \RRd}|x - y|^p \, \pi(dx, dy)\right)^{1/p} \in [0, \infty].
\]
Given $\pi \in \Pi(\mu,\nu)$, we write $\pi(dx,dy)=\mu(dx)\pi_x(dy)$ for its disintegration with respect to the first marginal. For $\mu,\nu\in\Prob_1(\RRd)$, we denote by $\Mart(\mu,\nu)$ the set of couplings $\pi \in \Pi(\mu,\nu)$ satisfying $\int_{\RRd}y\,\pi_x(dy)=x$ for $\mu$-a.e.\ $x$, and refer to its elements as martingale couplings from $\mu$ to $\nu$.

Let $T:\RRd\to\RRd$ be Borel measurable and $\xi\in\Prob(\RRd)$. The push-forward of $\xi$ under $T$ is denoted by $T_\#\xi$ and defined by $T_\#\xi(A):=\xi(T^{-1}(A))$ for every Borel set $A \subseteq \RRd$.

We denote by $\|\cdot\|_{\mathrm{HS}}$ the Hilbert--Schmidt norm on $\RRdd$. For a proper function $v:\RRd\to (-\infty, \infty]$, its convex conjugate is denoted by $v^*(y) := \sup_{x\in\RRd}{x\cdot y - v(x)}$.

\section{$q_x$-Bass martingales and the $q_x^{s,t}$-Bass extension}\label{sec:qx-Bass}

We begin this section by recalling the main definitions and results on $q_x$-Bass martingales from \cite{Ma26_qxBass}, and then specify the notation and assumptions for the continuous-time reference martingale used as our reference process. Its transition kernel from time $s$ to time $t$, denoted by $q_x^{s,t}$, is then used to introduce the notion of a $q_x^{s,t}$-Bass extension. As in the classical Bass construction, this extension is obtained by taking conditional expectations of a terminal random variable with respect to the filtration of the reference process. Although such an extension need not solve \eqref{eq:mainPb}, it provides the building block for constructing continuous-time processes that approximate an optimizer of that problem.

\begin{definition}[$q_x$-Bass martingale]
\label{def:qx-bass}
	Let $\mu,\nu\in\Prob_1(\RRd)$ and $(q_x)_{x\in\supp(\mu)} \subseteq \Prob(\RRd)$. We say that $\pi^*\in\Mart(\mu,\nu)$ is a $q_x$-Bass martingale from $\mu$ to $\nu$ if there exist a proper l.s.c. convex function $v:\RRd\to (-\infty, \infty]$ and a Borel function $a:\RRd\to\RRd$ such that, for $\mu$-a.e. $x$,
	\[
		v(a(x) + \cdot) \in L^1(q_x), \qquad \nabla v(a(x) + \cdot) \text{ exists } q_x\text{-a.s.},
	\]
	and, whenever $M_0 \sim \mu$ and $\Law(Q| M_0=x)=q_x$, one has
    \[
        \Law\bigl(M_0, \nabla v(a(M_0) + Q)\bigr) = \pi^*.
    \]
\end{definition}

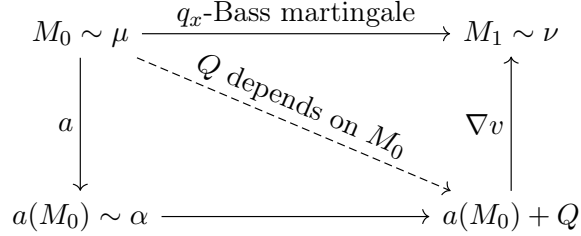
\begin{figure}[H]
\centering
\begin{tikzcd}[row sep=0.7in, column sep=1.4in]
  M_0 \sim \mu
  \arrow[r, "\text{\normalsize $q_x$-Bass martingale}"]
  \arrow[d, "\text{\normalsize $a$}"']
  \arrow[dr, dashed, "\text{\normalsize $Q$ depends on $M_0$}"{sloped}]
  &
  M_1 \sim \nu
  \\
  a(M_0) \sim \alpha
  \arrow[r]
  &
  a(M_0)+Q
  \arrow[u, "\text{\normalsize $\nabla v$}"]
\end{tikzcd}
\caption{Diagram for the $q_x$-Bass martingale from $\mu$ to $\nu$}
\label{fig:diagram}
\end{figure}

Figure \ref{fig:diagram} illustrates the construction of a $q_x$-Bass martingale, highlighting the roles of the maps $a$ and $\nabla v$ and the reference kernels $q_x$. For a detailed explanation of this construction in words, we refer to \cite[Section~2]{Ma26_qxBass}.

The theory developed in \cite{Ma26_qxBass} provides existence, optimality, and duality results for $q_x$-Bass martingales under suitable assumptions. In particular, if $\mu$ and $\nu$ are compactly supported, $\supp(\mu) \subseteq \interior(\co(\supp(\nu)))$, and the pair $(\mu,\nu)$ is irreducible, \cite[Theorem~4.3]{Ma26_qxBass} establishes the existence of a $q_x$-Bass martingale from $\mu$ to $\nu$. For the reader's convenience, we recall the definition of irreducibility:

\begin{definition}[Convex order and irreducibility]
\label{def:cvx-order-irreducibility}
Let $\mu,\nu\in\Prob_1(\RRd)$. We say that $\mu$ is dominated by $\nu$ in convex order, and write $\mu\preceq_c\nu$, if
\[
    \int_\RRd \varphi \, d\mu \leq \int_\RRd \varphi \, d\nu
\]
for every convex function $\varphi:\RRd \to \RR$ for which both integrals are well defined. The pair $(\mu,\nu)$ is called irreducible if $\mu\preceq_c\nu$ and, for every pair of Borel sets $A,B\subseteq\RRd$ with $\mu(A)>0$ and $\nu(B)>0$, there exists $\pi\in\Mart(\mu,\nu)$ such that
\begin{equation}
\label{eq:irreducible_coupling}
    \pi(A \times B) > 0.
\end{equation}
\end{definition}

Under suitable integrability conditions, a $q_x$-Bass martingale is also an optimizer of \eqref{eq:qx_WOT}; see \cite[Theorem~2.3]{Ma26_qxBass}.

Additionally, under the assumptions of \cite[Corollary~3.4]{Ma26_qxBass}, \eqref{eq:qx_WOT} is finite, the optimal martingale coupling is unique, and the problem admits the following dual representation
\begin{equation}\label{eq:dualPb}
    WT_{\Ical}^{q_\cdot}(\mu, \nu) = 2 \hspace{-5mm} \sup_{\substack{\psi\in L^1(\nu) \\ \psi \ \mathrm{proper\ l.s.c.}\\ \mathrm{convex}}} \hspace{-2mm} \int_{\RRd} \left[ \psi^{C_{q_\cdot}}(x) + \frac12 \int_{\RRd} |z|^2 \, q_x(dz) \right] \mu(dx) - \int_{\RRd} \left[ \psi(y) - \frac{|y|^2}{2} \right] \nu(dy),
\end{equation}
where the $C_{q_\cdot}$-transform of a proper l.s.c. convex function $\psi$ is defined by
\[
\psi^{C_{q_\cdot}}(x) := \sup_{a\in\RRd}\left(a\cdot x - \int_{\RRd}\psi^*(a + z)\,q_x(dz)\right).
\]

The dual formulation also provides an algorithm for computing a $q_x$-Bass martingale. More specifically, the Generalized Martingale Sinkhorn Algorithm introduced in \cite[Algorithm~1]{Ma26_qxBass} allows one to easily compute an optimal convex potential $\psi$ for \eqref{eq:dualPb}. Its convex conjugate $v=\psi^*$ provides a possible choice of the function $v$ in Definition~\ref{def:qx-bass}, from which the corresponding map $a$ can be recovered to construct the associated $q_x$-Bass martingale. 

\begin{algorithm}[H]
\caption{Generalized Martingale Sinkhorn algorithm}
\label{alg:q-sinkhorn}
\noindent Let $\psi_0 \in L^1(\nu)$ be a proper l.s.c. convex function such that $\psi_0^*$ is Lipschitz.

\noindent For $k\geq 1$:
\begin{enumerate}[label=(\roman*)]
    \item \label{it:q-sinkhorn-1} Define a Borel map $a_k:\supp(\mu) \to \RRd$ such that
    \[
         a_k(x) \in \operatorname*{arg \,max}_{a\in\RRd} \left(a \cdot x - \int_\RRd \psi_{k-1}^*(a + z) \,q_x(dz)\right).\vspace{-2mm}
    \]
    so that $\psi_{k-1}^{C_{q_\cdot}}(x) = a_k(x)\cdot x - \int_\RRd \psi_{k-1}^*(a_k(x) + z) \,q_x(dz)$.

    \item \label{it:q-sinkhorn-2} Define
    \[
        \rho_k := (a_k(x) + z)_\# \left(q_x(dz)\mu(dx)\right).
    \]

    \item \label{it:q-sinkhorn-3} Choose a proper l.s.c. convex function $\psi_k$ such that $\psi_k^*$ is a Brenier potential from $\rho_k$ to $\nu$, i.e.
    \[
        \nu = (\nabla\psi_k^*)_\#\rho_k.
    \]
\end{enumerate}
\end{algorithm}

Under the same compactness and irreducibility assumptions ensuring the existence of a $q_x$-Bass martingale from $\mu$ to $\nu$, suitably normalized subsequences of the convex potentials generated by Algorithm \ref{alg:q-sinkhorn} converge to potentials that generate $q_x$-Bass martingales \cite[Theorem~4.3]{Ma26_qxBass}. In dimension one, the algorithm reduces to a fixed-point iteration \cite[Section~5]{Ma26_qxBass}. We will also make use of the rescaling invariance established in \cite[Proposition~2.5]{Ma26_qxBass}: for any $s>0$, replacing $q_x$ with $(s\,\mathrm{Id})_\#q_x$ leaves the associated $q_x$-Bass martingale coupling unchanged.

We now introduce the continuous-time reference martingale underlying our calibration problem, whose transition kernels and filtration will then be used to extend the two-step $q_x$-Bass martingale construction to continuous time. 

Fix $0\leq t_0<t_1<\infty$ and a filtered probability space $(\Omega, \mathcal F, (\mathcal F_t)_{t\in[t_0,t_1]}, \PP)$ satisfying the usual conditions and rich enough to support a standard $d$-dimensional Brownian motion $(B_t)_{t \in [t_0, t_1]}$. We also assume that $\mathcal F_{t_0}$ supports a uniformly distributed random variable on $(0,1)$, so that any prescribed square-integrable initial distribution can be realized on this space.

Let $A \subseteq \RRd$ be an open convex set and $\widehat\sigma:[t_0, t_1] \times A \to \RRdd$. For $s\in[t_0,t_1]$ and any $A$-valued, square-integrable, $\mathcal F_s$-measurable random variable $\xi$, we denote by $\widehat M^{s,\xi}$ the solution of
\begin{equation}
\label{eq:reference_diffusion}
d\widehat M_t^{s,\xi} = \widehat\sigma(t, \widehat M_t^{s,\xi})\,dB_t, \qquad \widehat M_s^{s,\xi} = \xi, \qquad t \in [s, t_1].
\end{equation} 
We assume that \eqref{eq:reference_diffusion} admits a unique $A$-valued strong solution $\widehat M^{s,\xi}$ for every such $s$ and $\xi$, and that this solution is a martingale. When the initial condition is deterministic, $\xi=x\in A$, we simply write $\widehat M^{s,x}$. The transition kernels associated with this Markov family are given by
\[
q_x^{s,t} := \Law(\widehat M_t^{s,x}), \qquad t_0 \leq s < t \leq t_1, \quad x \in A.
\]
We also assume that $q_x^{s,t} \ll \lambda^d$, for all $s, t \in [t_0, t_1]$ and $x \in A$.

In addition to the above assumptions, the following conditions will be imposed as needed in the results throughout the paper:
\begin{assumption}[Further assumptions on the reference martingale]\label{ass:reference-martingale}
\noindent \begin{enumerate}[label=(A\arabic*)]
    \item \label{ass:A1} For every $s\in [t_0, t_1]$ and every square-integrable $A$-valued initial condition $\xi$, the solution $\widehat M^{s,\xi}$ is square-integrable.
    \item \label{ass:A2} There exists $L>0$ such that
    \[
        \|\widehat\sigma(t, x) - \widehat\sigma(t, y)\|_{HS} \leq L|x - y|, \qquad \text{for all $t\in [t_0, t_1]$ and $x,y\in A$}.
    \]
    \item \label{ass:A3} There exists $C_1>0$ such that
    \[
        |\widehat \sigma(t, x)^T \xi|^2 \geq C_1 |\xi|^2,
    \]
    for every $t \in [t_0, t_1]$, $x\in A$ and $\xi \in \RRd$.
    \item \label{ass:A4} There exists $C_2>0$ such that
    \[
        \|\widehat\sigma(t, x)\|_{HS} \leq C_2(1 + |x|),
    \]
    for every $t \in [t_0, t_1]$ and $x\in A$.
\end{enumerate}
\end{assumption}

The following proposition shows that the additional conditions~\ref{ass:A2}--\ref{ass:A4} imply our standing assumptions on the solutions of \eqref{eq:reference_diffusion}.

\begin{proposition}[Properties of the reference transition kernel]
\label{prop:reference-transition-kernel}
Suppose that \ref{ass:A2},~\ref{ass:A3} and~\ref{ass:A4} hold. Then, assuming that solutions of \eqref{eq:reference_diffusion} remain in $A$, \eqref{eq:reference_diffusion} admits a unique $A$-valued strong martingale solution for every square-integrable $A$-valued initial condition, and~\ref{ass:A1} holds. Moreover, for every $t_0\leq s<t\leq t_1$, the transition laws $(q_x^{s,t})_{x\in A}$ satisfy:
\begin{enumerate}[label=(\roman*)]
    \item \label{it:reference-transition-kernel-1} $q_x^{s,t}\ll \lambda^d$ for every $x\in A$;
    \item \label{it:reference-transition-kernel-2} $q_x^{s,t}\in\Prob_2(\RRd)$ for every $x\in A$;
    \item \label{it:reference-transition-kernel-3} the map $A\ni x\to q_x^{s,t}\in\Prob_2(\RRd)$ is Lipschitz continuous with respect to the $\Wcal_2$ distance.
\end{enumerate}
\end{proposition}

\begin{proof}
Fix $t_0\leq s<t\leq t_1$. Under \ref{ass:A2} and~\ref{ass:A4}, the Lipschitz and linear growth conditions ensure the existence and uniqueness of a strong solution of \eqref{eq:reference_diffusion}; see \cite[Theorem~5.2.9]{KaSh91}. Since the solutions are assumed to remain in $A$, they are well defined and $A$-valued throughout $[s,t_1]$ for every square-integrable $A$-valued initial condition. For \ref{it:reference-transition-kernel-1}, the Lipschitz, growth, and non-degeneracy conditions, together with the $A$-valued property of the strong solution, allow us to apply \cite[Theorem~2.3.1]{Nu}. Hence, $q_x^{s,t}\ll\lambda^d$ for every $x\in A$.

For \ref{it:reference-transition-kernel-2}, by Itô's isometry and~\ref{ass:A2}, for every $r\in[s,t]$,
\[
    \EE\left[|\widehat M_r^{s,x}|^2\right] \leq |x|^2 + 2C_2^2\int_s^r\left(1 + \EE\left[|\widehat M_u^{s,x}|^2\right]\right)du.
\]
Therefore, by Gronwall's inequality, $\EE[|\widehat M_t^{s,x}|^2]<\infty$, and hence $q_x^{s,t}\in\Prob_2(\RRd)$ for every $x\in A$.

To see \ref{it:reference-transition-kernel-3}, let $x,y\in A$ and consider the solutions $\widehat M^{s,x}$ and $\widehat M^{s,y}$ driven by the same Brownian motion. Then, for every $r\in[s,t]$, Itô's isometry and~\ref{ass:A2} imply
\[
    \EE\left[|\widehat M_r^{s,x} - \widehat M_r^{s,y}|^2\right] \leq |x - y|^2 + L^2\int_s^r\EE\left[|\widehat M_u^{s,x} - \widehat M_u^{s,y}|^2\right]du.
\]
Arguing as above, we obtain $\EE[|\widehat M_t^{s,x} - \widehat M_t^{s,y}|^2] \leq e^{L^2(t-s)}|x - y|^2$. Since the law of $(\widehat M_t^{s,x}, \widehat M_t^{s,y})$ is a coupling of $q_x^{s,t}$ and $q_y^{s,t}$, we have
\[
    0 \leq \Wcal_2(q_x^{s,t}, q_y^{s,t}) \leq \EE\left[|\widehat M_t^{s,x} - \widehat M_t^{s,y}|^2\right]^{1/2} \leq e^{\frac{L^2}{2}(t-s)}|x - y|.
\]
Therefore, the map $A\ni x\to q_x^{s,t}\in\Prob_2(\RRd)$ is Lipschitz continuous with respect to the $2$-Wasserstein distance.
\end{proof}

We now use the reference martingale to extend $q_x^{s,t}$-Bass martingale couplings to continuous time.

\begin{definition}[Continuous-time $q_x^{s,t}$-Bass extension]
Fix $t_0\leq s<t\leq t_1$ and $\mu,\nu\in\Prob_1(A)$ such that a $q_x^{s,t}$-Bass martingale from $\mu$ to $\nu$ exists. Let $\widehat M=(\widehat M_r)_{r\in[s,t]}$ be the reference martingale with initial law $\mu$, and denote by $\widehat\FF = (\widehat\Fcal_r)_{r\in[s,t]}$ its natural filtration.

Let $v$ and $a$ be the functions associated with the $q_x^{s,t}$-Bass martingale as in Definition~\ref{def:qx-bass}. The continuous-time $q_x^{s,t}$-Bass extension associated with $\widehat M$ is the $\widehat\FF$-martingale $(M_r)_{r\in[s,t]}$ defined by
\[
    M_r := \EE\left[ \nabla v\big(a(\widehat M_s) + \widehat M_t\big) \,\middle|\,\widehat\Fcal_r \right], \qquad r \in [s, t].
\]
\end{definition}
By construction, $\Law(M_s,M_t)$ is a $q_x^{s,t}$-Bass martingale coupling from $\mu$ to $\nu$.

This construction only requires the existence of a $q_x^{s,t}$-Bass coupling and suitable strong solutions of \eqref{eq:reference_diffusion}. Moreover, it admits the representation
\[
	M_r = M_s + \int_s^r\sigma^M_u\,dB_u, \qquad r \in [s, t],
\]
where $\sigma^M$ is predictable. In particular, under sufficient regularity, $\sigma^M_r$ can be expressed as a function of $(r,\widehat M_s,\widehat M_r)$.

\begin{remark}[On the (non-)optimality of the $q_x^{s,t}$-Bass extension associated with $\widehat M$]
\label{rmk:non_optimality}
Let $r\in[s,t]$. By the Markov property of $\widehat M$,
\[
    M_r = \int_{\RRd}\nabla v\big(a(\widehat M_s) + y\big)\,q_{\widehat M_r}^{r,t}(dy).
\]
In general, $\Law(M_s,M_r)$ is not a $q_x^{s,r}$-Bass martingale from $\Law(M_s)$ to $\Law(M_r)$.

This consistency property holds when the diffusion coefficient $\widehat\sigma$ is independent of the space variable. In this case, any solution of \eqref{eq:reference_diffusion} is a Gaussian additive process, and the continuous-time $q_x^{s,t}$-Bass extension associated with $\widehat M$ is an $\widehat M$-Bass martingale in the sense of \cite[Definition~7.2]{AcMa26_general}.
\end{remark}

\section{Approximation of dynamic martingale optimal transport}\label{sec:dynamic-approximation}

The $q_x^{s,t}$-Bass extension provides a way to pass from discrete martingales to continuous-time models while preserving the joint law of the endpoints, and hence the optimality of the corresponding coupling for \eqref{eq:qx_WOT}. As observed in Remark~\ref{rmk:non_optimality}, the joint laws at other pairs of times need not be optimal for the corresponding weak martingale transport problems. Nevertheless, we show in this section that this construction can be used to obtain continuous-time martingales which approximate the optimal value of
\[
	MOT(\Mart_{[t_0, t_1]}, \widehat \sigma) := \inf_{M \in \Mart_{[t_0,t_1]}} \EE\left[\int_{t_0}^{t_1}\|\sigma_t - \widehat\sigma(t, M_t)\|_{HS}^2\,dt\right].
\]

Here, $\Mart_{[t_0,t_1]}$ denotes the class of $A$-valued continuous martingales with square-integrable marginals, defined on the filtered probability space fixed in Section~\ref{sec:qx-Bass} and satisfying marginal constraints at finitely many maturities. These constraints may consist of fully prescribed marginal distributions or finitely many equalities or inequalities involving expectations of continuous payoff functions that are affine on finitely many convex regions covering $A$. The latter include, for instance, European calls and puts, basket options and spread options. The optimization is restricted to martingales admitting the representation
\[
M_t = M_{t_0} + \int_{t_0}^t \sigma_s\,dB_s, \qquad t \in [t_0, t_1],
\]
where $(\sigma_t)_{t \in [t_0, t_1]}$ is predictable and $(B_t)_{t \in [t_0, t_1]}$ is the fixed Brownian motion driving the reference diffusion in \eqref{eq:reference_diffusion}.
In particular, the constraints are chosen so that $\Mart_{[t_0,t_1]} \neq \emptyset$. For instance, one may prescribe the endpoint marginals $M_{t_0}\sim\mu$ and $M_{t_1}\sim\nu$ with $(\mu, \nu)$ irreducible, impose constraints of the form $\EE[(M_{s_i}-K_j)^+]=c_{i,j}$ for suitable $s_i\in[t_0,t_1]$, $K_j>0$, and $c_{i,j}>0$, or impose both types of constraints simultaneously, provided they are mutually consistent.

Let $n \in \NN$. Consider now a partition of $[t_0,t_1]$,
\[
    \Tcal := \{t_0 = s_1 < \dots < s_{n+1} = t_1\},
\]
such that all the times at which the constraints defining $\Mart_{[t_0,t_1]}$ are imposed belong to $\Tcal$. We define the discrete-time analogue of $\Mart_{[t_0,t_1]}$ by
\begin{align*}
    \Mcal_\Tcal := \Big\{(\eta_i)_{i=1}^{n+1} \in ( & \Prob_2(A))^{n+1}:\
    \eta_i \preceq_c \eta_{i+1}\text{ for all }i = 1, \dots, n, \\
    & \text{and, if any constraints are prescribed for }\Law(M_{s_i}), \ \eta_i\text{ satisfies them}
    \Big\}.
\end{align*}
Observe that a family $(\eta_i)_{i=1}^{n+1}\in\Mcal_\Tcal$ does not, in general, determine a unique discrete-time martingale. Indeed, there may exist several $(n+1)$-step martingales $(M_i)_{i=1}^{n+1}$ such that $\Law(M_i)=\eta_i$ for every $i=1,\dots,n+1$. Thus, while an element of $\Mart_{[t_0,t_1]}$ specifies a Brownian martingale on the whole time interval, an element of $\Mcal_\Tcal$ specifies only its prescribed one-time marginals on the grid $\Tcal$.

For this reason, we introduce a selection criterion based on $q_x$-Bass martingale transport. For each $i=1,\dots,n$, let $q_x^{(i)}:=q_x^{s_i,s_{i+1}}$ denote the transition kernel of the reference martingale over $[s_i,s_{i+1}]$. We then consider the following multimarginal weak martingale optimal transport problem
\[
    MWT \big(\Mcal_\Tcal, (q_\cdot^{(i)})_{i=1}^n\big) := \inf_{(\eta_i)_{i=1}^{n+1}\in\Mcal_\Tcal}\sum_{i=1}^n WT_{\Ical}^{q_\cdot^{(i)}}(\eta_i, \eta_{i+1}),
\]
where $WT_{\Ical}^{q_\cdot}$ is defined in \eqref{eq:qx_WOT}. The properties of the optimizers of \eqref{eq:WOT} can then be derived from those of the underlying $q_x$-Bass transport problems, recalled in Section~\ref{sec:qx-Bass}.

\begin{proposition}
\label{prop:approximate-problem}
Let $\Tcal=\{t_0=s_1<\dots<s_{n+1}=t_1\}$ and assume that $\Mcal_\Tcal\neq\emptyset$. Suppose that \ref{ass:A1} and~\ref{ass:A2} hold. Then the following statements hold:
\begin{enumerate}[label=(\roman*)]
    \item \label{it:approximate-problem-1} For every $(\eta_i)_{i=1}^{n+1}\in\Mcal_\Tcal$,
    \[
    \sum_{i=1}^n WT_{\Ical}^{q_\cdot^{(i)}}(\eta_i, \eta_{i+1}) < \infty.
    \]
    In particular, $MWT(\Mcal_\Tcal, (q_\cdot^{(i)})_{i=1}^n) < \infty$.

    \item \label{it:approximate-problem-2} For any $(\eta_i)_{i=1}^{n+1}\in\Mcal_\Tcal$, there exists a unique Markov martingale measure on $(\RRd)^{n+1}$ with marginals $(\eta_i)_{i=1}^{n+1}$ whose consecutive two-time marginals are optimizers of
    \[
    WT_{\Ical}^{q_\cdot^{(i)}}(\eta_i, \eta_{i+1}), \qquad i = 1, \dots, n.
    \]
    It is given by the Markovian gluing of the unique optimizers of $WT_{\Ical}^{q_\cdot^{(i)}}(\eta_i, \eta_{i+1})$, $i=1,\dots,n$.

    \item \label{it:approximate-problem-3} The map
    \[
    (\eta_i)_{i=1}^{n+1} \longmapsto \sum_{i=1}^n WT_{\Ical}^{q_\cdot^{(i)}}(\eta_i, \eta_{i+1})
    \]
    is convex on $\Mcal_\Tcal$. In particular, \eqref{eq:WOT} is a convex optimization problem.
\end{enumerate}
\end{proposition}

\begin{proof}
It follows from \ref{ass:A1} that, for every $(\eta_i)_{i=1}^{n+1}\in\Mcal_\Tcal$,
\[
\int_{\RRd\times\RRd}|y|^2\,q_x^{(i)}(dy)\,\eta_i(dx) < \infty,
\]
since $\eta_i\in\Prob_2(A)$. Therefore, \cite[Corollary~3.4]{Ma26_qxBass} yields $WT_{\Ical}^{q_\cdot^{(i)}}(\eta_i, \eta_{i+1}) < \infty$ for all $i=1,\dots,n$, which proves \ref{it:approximate-problem-1}.

To see \ref{it:approximate-problem-2}, \cite[Corollary~3.4]{Ma26_qxBass} yields also, for every $i=1,\dots,n$, a unique optimizer $\pi^{(i)} \in \Mart(\eta_i, \eta_{i+1})$. Their Markovian gluing defines the unique Markov martingale measure with marginals $(\eta_i)_{i=1}^{n+1}$ and consecutive two-time marginals $(\pi^{(i)})_{i=1}^n$.

Finally, for \ref{it:approximate-problem-3}, the dual formulation in \eqref{eq:dualPb} shows that $(\rho, \zeta) \mapsto WT_{\Ical}^{q_\cdot^{(i)}}(\rho, \zeta)$ is convex, since it is the supremum of linear functionals in $(\rho,\zeta)$. As $\Mcal_\Tcal$ is a convex set, the map
\[
(\eta_i)_{i=1}^{n+1} \longmapsto \sum_{i=1}^n WT_{\Ical}^{q_\cdot^{(i)}}(\eta_i, \eta_{i+1})
\]
is convex on $\Mcal_\Tcal$.
\end{proof}

Building on the notion of $q_x^{s,t}$-Bass extension, we now show how to construct a continuous-time martingale model from an element of $\Mcal_\Tcal$.

\begin{definition}[Continuous-time $(q_x^{(i)})_{i=1}^n$-Bass interpolation]
\label{def:continuous-time-q-Bass-interpolation}
Let $\Tcal=\{t_0=s_1<\dots<s_{n+1}=t_1\}$ and let $\eta = (\eta_i)_{i=1}^{n+1} \in \Mcal_\Tcal$ be such that, for every $i=1,\dots,n$, there exists a $q_x^{(i)}$-Bass martingale from $\eta_i$ to $\eta_{i+1}$.

For every $i=1,\dots,n$, let $\widehat M^{(i)}$ be a solution of \eqref{eq:reference_diffusion} on $[s_i,s_{i+1}]$ with $\Law(\widehat M_{s_i}^{(i)})=\eta_i$, and let $M^{(i)}$ be the corresponding continuous-time $q_x^{(i)}$-Bass extension. We define the continuous-time $(q_x^{(i)})_{i=1}^n$-Bass interpolation of $(\eta_i)_{i=1}^{n+1}$ as the martingale $(M_t^\eta)_{t\in[t_0,t_1]}$ obtained by gluing $(M^{(i)})_{i=1}^n$ along the marginals $(\eta_i)_{i=1}^{n+1}$, namely
\[
M_t^\eta = M_t^{(i)}, \qquad t \in [s_i, s_{i+1}], \quad i = 1, \dots, n,
\]
where the processes are glued so that $\Law(M_{s_i}^\eta)=\eta_i$, for all $i=1,\dots,n+1$. 
\end{definition}

By construction, the consecutive two-time marginals of $M^\eta$ are the $q_x^{(i)}$-Bass martingales from $\eta_i$ to $\eta_{i+1}$. In particular, $\Law(M_{s_i}^\eta,M_{s_{i+1}}^\eta)$ is an optimizer of $WT_{\Ical}^{q_\cdot^{(i)}}(\eta_i, \eta_{i+1})$ for every $i=1,\dots,n$. Moreover, $M^\eta$ satisfies all the marginal constraints prescribed on the grid $\Tcal$.

We are now in a position to establish the connection between \eqref{eq:mainPb} and~\eqref{eq:WOT}.

\begin{theorem}[Approximation of the value of \eqref{eq:mainPb}]
\label{thm:approximate-dynamic-MOT}
Let $\Tcal=\{t_0=s_1<\dots<s_{n+1}=t_1\}$. Suppose that \ref{ass:A1} and~\ref{ass:A2} hold. Assume that $(\eta_i^*)_{i=1}^{n+1} \in \Mcal_\Tcal$ is an optimizer of \eqref{eq:WOT} admitting a continuous-time $(q_x^{(i)})_{i=1}^n$-Bass interpolation $M^{\eta^*}$ of $(\eta_i^*)_{i=1}^{n+1}$ with Brownian representation  $dM_t^{\eta^*} = \sigma_t^{\eta^*}\,dB_t$. Set
\[
|\Tcal| := \max_{i=1,\dots,n}(s_{i+1} - s_i).
\]
Then there exists a constant $C>0$, independent of $\Tcal$, such that
\[
0 \leq \EE\left[\int_{t_0}^{t_1}\|\sigma_t^{\eta^*} - \widehat\sigma(t, M_t^{\eta^*})\|_{HS}^2\,dt\right] - MOT(\Mart_{[t_0,t_1]}, \widehat\sigma) \leq C\sqrt{|\Tcal|}.
\]
\end{theorem}

\begin{proof}
Set $h_i:=s_{i+1}-s_i$, for $i=1,\dots,n$ and denote by $M^{(i)}$ the continuous-time $q_x^{(i)}$-Bass extension associated with $M^{\eta^*}$. Applying the second estimate in Lemma~\ref{lem:one-step-estimate} to each $M^{(i)}$, we obtain
\[
\EE\left[\int_{s_i}^{s_{i+1}}\|\sigma_r^{(i)} - \widehat\sigma(r, M_r^{(i)})\|_{HS}^2\,dr\right] \leq (1 + C\sqrt{h_i})WT_{\Ical}^{q_\cdot^{(i)}}(\eta_i^*, \eta_{i+1}^*).
\]
Since $h_i\leq|\Tcal|$, summing over $i=1,\dots,n$ yields
\begin{equation}
\label{eq:estimate1_mainApproxThm}
\EE\left[\int_{t_0}^{t_1}\|\sigma_r^{\eta^*} - \widehat\sigma(r, M_r^{\eta^*})\|_{HS}^2\,dr\right] \leq (1 + C\sqrt{|\Tcal|})\sum_{i=1}^n WT_{\Ical}^{q_\cdot^{(i)}}(\eta_i^*, \eta_{i+1}^*).
\end{equation}

Let now $M \in\Mart_{[t_0,t_1]}$ with representation $dM_t = \sigma_t dB_t$ and set $\eta_i:=\Law(M_{s_i})$. Since $\Tcal$ contains all the constraint times, $(\eta_i)_{i=1}^{n+1}\in\Mcal_\Tcal$. By the optimality of $(\eta_i^*)_{i=1}^{n+1}$,
\begin{equation}
\label{eq:estimate2_mainApproxThm}
\sum_{i=1}^n WT_{\Ical}^{q_\cdot^{(i)}}(\eta_i^*, \eta_{i+1}^*) \leq \sum_{i=1}^n WT_{\Ical}^{q_\cdot^{(i)}}(\eta_i, \eta_{i+1}).
\end{equation}
Applying the first estimate in Lemma~\ref{lem:one-step-estimate} to the restriction of $M$ to each interval $[s_i,s_{i+1}]$, we have
\[
WT_{\Ical}^{q_\cdot^{(i)}}(\eta_i, \eta_{i+1}) \leq (1 + C\sqrt{h_i})\EE\left[\int_{s_i}^{s_{i+1}}\|\sigma_r - \widehat\sigma(r, M_r)\|_{HS}^2\,dr\right].
\]
Hence,
\begin{equation}
\label{eq:estimate3_mainApproxThm}
\sum_{i=1}^n\ WT_{\Ical}^{q_\cdot^{(i)}}(\eta_i, \eta_{i+1}) \leq (1 + C\sqrt{|\Tcal|})\EE\left[\int_{t_0}^{t_1}\|\sigma_r - \widehat\sigma(r, M_r)\|_{HS}^2\,dr\right].
\end{equation}
Combining \eqref{eq:estimate1_mainApproxThm},~\eqref{eq:estimate2_mainApproxThm} and~\eqref{eq:estimate3_mainApproxThm} and taking the infimum over $M \in\Mart_{[t_0,t_1]}$ gives
\[
\EE\left[\int_{t_0}^{t_1}\|\sigma_r^{\eta^*} - \widehat\sigma(r, M_r^{\eta^*})\|_{HS}^2\,dr\right] \leq (1 + C\sqrt{|\Tcal|})^2MOT(\Mart_{[t_0,t_1]}, \widehat\sigma).
\]
Since $|\Tcal|\leq t_1-t_0$,
\[
(1 + C\sqrt{|\Tcal|})^2 \leq 1 + C'\sqrt{|\Tcal|},
\]
where $C':=2C+C^2\sqrt{t_1-t_0}$. It follows that
\[
0 \leq \EE\left[\int_{t_0}^{t_1}\|\sigma_r^{\eta^*} - \widehat\sigma(r, M_r^{\eta^*})\|_{HS}^2\,dr\right] - MOT(\Mart_{[t_0,t_1]}, \widehat\sigma) \leq C'MOT(\Mart_{[t_0,t_1]}, \widehat\sigma)\sqrt{|\Tcal|}.
\]
Absorbing the fixed quantity $C'MOT(\Mart_{[t_0,t_1]}, \widehat\sigma)$ into the constant concludes the proof.
\end{proof}

\section{Numerical methods}\label{sec:numerical-algorithm}
In this section, we introduce methods for computing an optimizer of \eqref{eq:WOT} that are based on the dual formulation \eqref{eq:dualPb}. Indeed, \eqref{eq:WOT} can be written as
\[
MWT\big(\Mcal_\Tcal, (q_\cdot^{(i)})_{i=1}^n\big) = \inf_{\eta \in\Mcal_\Tcal}F(\eta),
\]
where $F:\Prob_2(A)^{n+1}\to\RR$ is given by
\begin{equation}
\label{eq:inf_sup_WOT}
\begin{aligned}
F(\eta) := 2 \sum_{i=1}^n\sup_{\substack{\psi_i\in L^1(\eta_{i+1})\\ \psi_i\ \mathrm{proper}\\ \mathrm{l.s.c.\ convex}}}
\Bigg( \int_{\RRd\times\RRd}\left(\psi_i^{C_{q_\cdot^{(i)}}}(x) + \frac{|y|^2}{2}\right) & q_x^{(i)}(dy)\,\eta_i(dx) \\
& - \int_{\RRd}\left(\psi_i(y) - \frac{|y|^2}{2}\right)\eta_{i+1}(dy) \Bigg).
\end{aligned}
\end{equation}

The additive structure of $F$ has two useful computational consequences: the dual potentials can be computed in parallel (Remark~\ref{rmk:parallel-computation}), and fully prescribed intermediate marginals split the problem into independent subproblems (Proposition~\ref{prop:decomposition-fixed-marginals}).

\begin{remark}[Parallel computation of the subgradient's potentials]\label{rmk:parallel-computation}
Once the current chain of marginals $\eta=(\eta_i)_{i=1}^{n+1}$ is fixed, the $n$ one-step problems
\[
    WT_{\Ical}^{q_\cdot^{(i)}}(\eta_i, \eta_{i+1}),  \qquad i = 1, \dots,  n,
\]
are independent. Hence, although the original optimization problem involves a chain of $n+1$ marginals, the computation of the corresponding dual potentials reduces to solving in parallel $n$ two-marginal problems by means of Algorithm~\ref{alg:q-sinkhorn}.
\end{remark}

\begin{proposition}[Decomposition along prescribed marginals]
\label{prop:decomposition-fixed-marginals}
Let $\eta = (\eta_i)_{i=1}^{n+1} \in \Mcal_\Tcal$ and suppose that, for some $j\in\{2,\dots,n\}$, the marginal $\eta_j$ is fully prescribed. Then
\[
\sum_{i=1}^nWT_{\Ical}^{q_\cdot^{(i)}}(\eta_i, \eta_{i+1}) = \sum_{i=1}^{j-1}WT_{\Ical}^{q_\cdot^{(i)}}(\eta_i, \eta_{i+1}) + \sum_{i=j}^nWT_{\Ical}^{q_\cdot^{(i)}}(\eta_i, \eta_{i+1}),
\]
and minimizing \eqref{eq:WOT} is equivalent to minimizing separately the two terms on the right-hand side over the corresponding admissible subchains, with $\eta_j$ kept fixed.
\end{proposition}

Furthermore, the following result establishes that the dual representation \eqref{eq:inf_sup_WOT} provides, in addition to the value of $F$, a natural linearization of $F$ with respect to the marginals chain in $\Mcal_\Tcal$.
\begin{proposition}
\label{prop:subgradient-F}
Let $\widehat \eta = (\widehat \eta_i)_{i=1}^{n+1} \in \Mcal_\Tcal$ admit a continuous-time $(q_x^{(i)})_{i=1}^n$-Bass interpolation as in Definition~\ref{def:continuous-time-q-Bass-interpolation}, and let $(\widehat \psi_i)_{i=1}^n$ be the corresponding optimal dual potentials. Define
\[
\begin{aligned}
\Lcal(\eta, \widehat \psi) := 2\sum_{i=1}^n\Bigg(\int_{\RRd\times\RRd}\left((\widehat \psi_i)^{C_{q_\cdot^{(i)}}}(x) + \frac{|y|^2}{2}\right) & q_x^{(i)}(dy)\,\eta_i(dx)\\
& - \int_{\RRd}\left(\widehat \psi_i(y) - \frac{|y|^2}{2}\right)\eta_{i+1}(dy)\Bigg).
\end{aligned}
\]
Then $F(\widehat \eta) = \Lcal(\widehat \eta, \widehat \psi)$ and $F(\eta)\geq\Lcal(\eta,\widehat \psi)$ for every $\eta$ such that the integrals in $\Lcal(\eta,\widehat \psi)$ are well-defined. In particular,
\[
F(\eta) \geq F(\widehat \eta) + \Lcal(\eta - \widehat \eta, \widehat \psi),
\]
and $\Lcal(\cdot,\widehat \psi)$ determines a subgradient of $F$ at $\widehat \eta$.
\end{proposition}

\begin{proof}
By the dual representation \eqref{eq:inf_sup_WOT}, $F(\eta)$ is obtained by taking the supremum over all admissible families of dual potentials. Therefore, evaluating the dual objective at the fixed potentials $\widehat\psi = (\widehat\psi_i)_{i=1}^n$ gives $F(\eta)\geq\Lcal(\eta,\widehat\psi)$, provided these potentials are admissible for $\eta$. On the other hand, since $\widehat\psi_i$ is optimal for the one-step problem associated with $(\widehat\eta_i,\widehat\eta_{i+1})$, we have $F(\widehat\eta) = \Lcal(\widehat\eta, \widehat\psi)$.
Combining these two relations and using the linearity of $\Lcal$ in its first argument yields
\[
F(\eta) - F(\widehat\eta) \geq \Lcal(\eta, \widehat\psi) - \Lcal(\widehat\eta, \widehat\psi) = \Lcal(\eta - \widehat\eta, \widehat\psi),
\]
which proves the claimed subgradient inequality.
\end{proof}

We now introduce the finite-dimensional setting in which the minimization algorithms for the discretization of \eqref{eq:WOT} will be formulated. For the sake of clarity, in view of Proposition~\ref{prop:decomposition-fixed-marginals}, we restrict the discussion to the case in which no intermediate marginal is fully prescribed. Otherwise, the problem decomposes into independent subproblems of the same form. Likewise, we may assume that neither the initial nor the terminal marginal is fully prescribed. If one of them is fixed, it is not part of the optimization procedure and serves only as a boundary condition on the admissible chain. Finally, to obtain a finite-dimensional formulation and facilitate the convergence analysis, we discretize the admissible marginals on finite grids, possibly different at each time step.

Fix $m \in \NN$ and
$\widehat X \in A^{(n+1)\times m}$
such that
\begin{equation}
\label{eq:support_irreducible}
\widehat x_{ij} \in \interior\left( \co\{\widehat x_{(i+1)k}:k = 1, \dots, m\} \right), \qquad i = 1, \dots, n, \quad j = 1, \dots, m,
\end{equation}
and define
\[
\Mcal_{\Tcal,\widehat X} := \left\{ (\eta_i)_{i=1}^{n+1} \in \Mcal_\Tcal : \eta_i = \sum_{j=1}^m p_{ij} \delta_{\widehat x_{ij}}, \ p_{ij} \in [0, 1], \ \sum_{j=1}^m p_{ij} = 1, \ i = 1, \dots, n + 1 \right\}.
\]
Each element of $\Mcal_{\Tcal,\widehat X}$ can be identified with a row-stochastic matrix $P=(p_{ij})\in[0,1]^{(n+1)\times m}$. We denote by $K_{\Tcal,\widehat X} \subseteq \RR^{(n+1)\times m}$ the corresponding set of admissible matrices, obtained by translating the convex-order and marginal constraints defining $\Mcal_\Tcal$ into constraints on the entries of $P$. In particular, $K_{\Tcal,\widehat X}$ is a finite-dimensional convex set.

For $P\in K_{\Tcal,\widehat X}$ and $i=1,\dots,n+1$, let
\[
\eta_i(P) := \sum_{j=1}^m p_{ij}\delta_{\widehat x_{ij}},
\qquad
\eta(P) := (\eta_i(P))_{i=1}^{n+1},
\]
and define $\widetilde F:K_{\Tcal,\widehat X} \to \RR$ by $\widetilde F(P):=F(\eta(P))$. The discretization of \eqref{eq:WOT} can then be written as
\begin{equation}
\label{eq:discretized_WOT}
\inf_{P\in K_{\Tcal,\widehat X}}\widetilde F(P).
\end{equation}
Its consistency with the original problem is addressed by the following result, whose proof is postponed to the appendix.

For $m\in\NN$, let $\mathfrak X_m$ denote the set of matrices $\widehat X = (\widehat x_{ij}) \in A^{(n+1)\times m}$ such that \eqref{eq:support_irreducible} holds.
\begin{proposition}[Stability of the discretization]
\label{prop:finite-support-approximation}
Let $\Tcal=\{t_0=s_1<\dots<s_{n+1}=t_1\}$ and assume that $\Mcal_\Tcal\neq\emptyset$. Suppose that \ref{ass:A1} and~\ref{ass:A2} hold. Then
\[
MWT\big(\Mcal_\Tcal, (q_\cdot^{(i)})_{i=1}^n\big) = \inf_{\substack{m\in\NN\\ \widehat X\in\mathfrak X_m}}\ \inf_{\eta\in\Mcal_{\Tcal,\widehat X}}F(\eta).
\]
\end{proposition}

Furthermore, we assume that there exists $\bar P\in\ri(K_{\Tcal,\widehat X})$ such that
\begin{equation}
\label{eq:chain-irreducibility}
(\eta_i(\bar P), \eta_{i+1}(\bar P)) \quad \text{is an irreducible pair for every $i=1,\dots,n$},
\end{equation}
and
\begin{equation}
\label{eq:chain-support}
\supp(\eta_i(\bar P)) \subseteq \interior\left(\co(\supp(\eta_{i+1}(\bar P)))\right), \qquad i = 1, \dots, n.
\end{equation}
To ensure that the chains considered in the minimization algorithms admit a continuous-time $(q_x^{(i)})_{i=1}^n$-Bass interpolation, we restrict the feasible set to a subset of $K_{\Tcal,\widehat X}$ bounded away from its relative boundary. For $\varepsilon>0$, define
\[
K_{\Tcal,\widehat X}^\varepsilon := \left\{P \in K_{\Tcal,\widehat X}:B_\varepsilon(P)\cap\aff(K_{\Tcal,\widehat X}) \subseteq K_{\Tcal,\widehat X}\right\}.
\]

As shown in the following proposition, the irreducibility and support conditions \eqref{eq:chain-irreducibility} and~\eqref{eq:chain-support} hold for every $P\in K_{\Tcal,\widehat X}^\varepsilon$. These conditions ensure the existence of a $q_x^{(i)}$-Bass martingale between $\eta_i(P)$ and $\eta_{i+1}(P)$ for each $i=1,\dots,n$, and hence of a continuous-time $(q_x^{(i)})_{i=1}^n$-Bass interpolation for the chain.

\begin{proposition}[Properties of $K_{\Tcal,\widehat X}^\varepsilon$]
\label{prop:properties-K-epsilon}
For every $\varepsilon>0$, the set $K_{\Tcal,\widehat X}^\varepsilon$ is convex. Moreover, \eqref{eq:chain-irreducibility} and~\eqref{eq:chain-support} hold for every $P\in K_{\Tcal,\widehat X}^\varepsilon$ in place of $\bar P$. In particular, for every $P\in K_{\Tcal,\widehat X}^\varepsilon$, there exists a continuous-time $(q_x^{(i)})_{i=1}^n$-Bass interpolation for $\eta(P)$.
\end{proposition}

\begin{proof}
The convexity of $K_{\Tcal,\widehat X}^\varepsilon$ follows from that of $K_{\Tcal,\widehat X}$. Fix $P \in K_{\Tcal,\widehat X}^\varepsilon \subseteq \ri(K_{\Tcal,\widehat X})$. Since $P$ and $\bar P$ belong to the relative interior, their marginal supports coincide, so \eqref{eq:chain-support} holds for $P$. Moreover, there exist $\lambda\in(0,1)$ and $Q\in K_{\Tcal,\widehat X}$ such that $P=\lambda\bar P+(1-\lambda)Q$. For every $i=1,\dots,n$ and Borel sets $A,B$ with $\eta_i(P)(A)>0$ and $\eta_{i+1}(P)(B)>0$, by \eqref{eq:irreducible_coupling} there exists a martingale coupling $\bar\pi$ of $(\eta_i(\bar P),\eta_{i+1}(\bar P))$ such that $\bar\pi(A\times B)>0$. Hence $\lambda\bar\pi+(1-\lambda)\pi_Q$, where $\pi_Q$ is any martingale coupling of $(\eta_i(Q),\eta_{i+1}(Q))$, is a martingale coupling of $(\eta_i(P),\eta_{i+1}(P))$ assigning positive mass to $A\times B$. Thus \eqref{eq:chain-irreducibility} holds for $P$.

Finally, the existence of a $q_\cdot^{(i)}$-Bass martingale from $\eta_i(P)$ to $\eta_{i+1}(P)$, follows from \cite[Theorem~4.3]{Ma26_qxBass}.
\end{proof}

The results above suggest a preliminary approach to optimize
\eqref{eq:WOT} based on standard constrained optimization methods, using the optimal dual potentials to evaluate $\widetilde F$ and its subgradients in parallel.

\begin{algorithm}[H]
\caption{Preliminary optimization algorithm}
\label{alg:preliminary-algorithm}
\noindent Choose $P^{(0)} \in K_{\Tcal,\widehat X}^\varepsilon$ and a numerical method for constrained convex optimization.\\
\noindent At each evaluation required by the optimization method:
\begin{enumerate}[label=(\roman*)]
    \item \label{step:preliminary-1} Given $P\in K_{\Tcal,\widehat X}^\varepsilon$, compute in parallel the optimal dual potentials $(\psi_i^P)_{i=1}^n$ associated with $WT_{\Ical}^{q_\cdot^{(i)}}(\eta_i(P), \eta_{i+1}(P))$, $i=1,\dots,n$, by means of Algorithm~\ref{alg:q-sinkhorn}.
    \item \label{step:preliminary-2} Use these potentials to evaluate $\widetilde F(P)$ and a subgradient $G(P)\in\partial\widetilde F(P)$, as in Proposition~\ref{prop:subgradient-F}.
    \item \label{step:preliminary-3} Use $\widetilde F(P)$ and $G(P)$ to update jointly the marginal weights through the chosen optimization method, subject to $P\in K_{\Tcal,\widehat X}^\varepsilon$.
\end{enumerate}
\noindent Repeat until a prescribed stopping criterion is satisfied.
\end{algorithm}

\begin{remark}
As $|\Tcal|\to0$, the kernels $q_x^{(i)}$ become increasingly concentrated around $x$, which may create numerical difficulties in step~\ref{step:preliminary-1} of Algorithm~\ref{alg:preliminary-algorithm}. This can be mitigated using \cite[Proposition~2.5]{Ma26_qxBass}. Indeed, for $s > 0$, replacing $q_x^{(i)}$ by $\widetilde q_x^{(i)} := (s \,\mathrm{Id})_\#q_x^{(i)}$ leaves the corresponding $q_x$-Bass martingale unchanged, while the associated dual potentials are rescaled accordingly. Thus, a suitable rescaling of the reference kernels can improve the numerical conditioning of Algorithm \ref{alg:q-sinkhorn}. However, since rescaling changes the value of each weak transport problem \eqref{eq:qx_WOT}, the potentials $\widetilde\psi_i^P$ computed with the rescaled kernels in step~\ref{step:preliminary-1} must be divided by $s$ before being used to compute the subgradient of the original objective in step~\ref{step:preliminary-1}, i.e., $\psi_i^P=\widetilde\psi_i^P/s$.
\end{remark}

This preliminary algorithm can be implemented using the wide range of numerical solvers available for constrained convex optimization. However, its generality makes it difficult to establish convergence results without specifying the optimization method. For this reason, we now focus on projected subgradient methods, whose convergence can be easily analyzed within our framework; see, for instance, \cite[Section~6.3]{Bertsekas09}. At each iteration, a subgradient of $\widetilde F$ is computed from the optimal dual potentials and used to update the marginal weights through an orthogonal projection onto $K_{\Tcal,\widehat X}^\varepsilon$. We choose positive step sizes $(\tau_k)_{k\in\NN}$ satisfying the classical convergence condition~\eqref{eq:convergence-condition}; see \cite[Theorem~1]{AlIuSo98}. A simple, though not necessarily the most effective, choice is $\tau_k=1/(k+1)$. This leads to the following algorithm.

\begin{algorithm}[H]
\caption{Projected Subgradient Minimization Algorithm}
\label{alg:projected-subgradient}
\noindent Choose $P^{(0)} \in K_{\Tcal,\widehat X}^\varepsilon$ and a sequence $(\tau_k)_{k\in\NN}$ of positive step sizes satisfying \eqref{eq:convergence-condition}.\\

\noindent For $k\in\NN$:
\begin{enumerate}[label=(\roman*)]
    \item \label{step:projected-subgradient-1} For each $i=1,\dots,n$, compute (in parallel) the convex potential $\psi_i^{(k)}$ associated with $WT_\Ical^{q_\cdot^{(i)}}(\eta_i(P^{(k)}), \eta_{i+1}(P^{(k)}))$ using Algorithm~\ref{alg:q-sinkhorn}.

    \item \label{step:projected-subgradient-2} Let $G^{(k)}$ be a subgradient of $\widetilde F$ at $P^{(k)}$ induced by $(\psi_i^{(k)})_{i=1}^n$, namely
    \[
    \langle G^{(k)}, P - P^{(k)}\rangle_{HS} = \Lcal(\eta(P) - \eta(P^{(k)}), \psi^{(k)}), \qquad P \in K_{\Tcal,\widehat X}.
    \]

    \item \label{step:projected-subgradient-3} Update
    \[
    P^{(k+1)} \in \operatorname*{argmin}_{P\in K_{\Tcal,\widehat X}^\varepsilon}\left\{\langle G^{(k)}, P\rangle_{HS} + \frac{1}{2\tau_k}\|P - P^{(k)}\|^2_{HS}\right\}.
    \]
\end{enumerate}
\end{algorithm}

The minimization in step~\ref{step:projected-subgradient-3} is equivalent to
\[
P^{(k+1)} = \Pi_{K_{\Tcal,\widehat X}^\varepsilon}\left(P^{(k)} - \tau_kG^{(k)}\right),
\]
where $\Pi_{K_{\Tcal,\widehat X}^\varepsilon}$ denotes the orthogonal projection onto $K_{\Tcal,\widehat X}^\varepsilon$ with respect to the Hilbert--Schmidt norm. Moreover, restricting the iterates to $K_{\Tcal,\widehat X}^\varepsilon\Subset\ri(K_{\Tcal,\widehat X})$ ensures that the subgradients of $\widetilde F$ are uniformly bounded; see \cite[Theorems~10.4 and~24.7]{Ro70}. Together with \eqref{eq:convergence-condition}, this yields the convergence in Theorem~\ref{thm:convergence-projected-subgradient}.

\begin{remark}[Approximate dual potentials]
Algorithm~\ref{alg:q-sinkhorn}, used in step~\ref{it:approximate-problem-1} of Algorithm~\ref{alg:projected-subgradient}, need not produce an optimal dual potential after finitely many iterations. Nevertheless, each iteration yields a strict improvement of the dual objective unless convergence has already been achieved; see \cite[Theorem~3.5]{Ma26_qxBass}. More generally, $\delta_k$-suboptimal dual potentials yield $\delta_k$-subgradients of the objective, and convergence results for projected subgradient methods with approximate subgradients are well established; see, e.g., \cite[Section~6.3.2]{Bertsekas09}. For simplicity, we restrict our convergence analysis to the idealized setting in which the dual potentials are computed exactly.
\end{remark}

\begin{theorem}[Convergence of Algorithm~\ref{alg:projected-subgradient}]
\label{thm:convergence-projected-subgradient}
Suppose that $\widetilde F$ is finite on $K_{\Tcal,\widehat X}$ and there exists $\varepsilon>0$ such that \eqref{eq:discretized_WOT} admits an optimizer $P^* \in K_{\Tcal,\widehat X}^\varepsilon\Subset\ri(K_{\Tcal,\widehat X})$. Let $(P^{(k)})_{k\in\NN}$ be generated by Algorithm~\ref{alg:projected-subgradient}, with step sizes satisfying
\begin{equation}
\label{eq:convergence-condition}
\tau_k > 0, \qquad \sum_{k=0}^\infty\tau_k = \infty, \qquad \sum_{k=0}^\infty\tau_k^2 < \infty.
\end{equation}
Then $(P^{(k)})_{k\in\NN}$ converges to an optimizer $P^\star \in K_{\Tcal,\widehat X}^\varepsilon$ of \eqref{eq:discretized_WOT}. In particular,
\[
\widetilde F(P^{(k)}) \longrightarrow \widetilde F(P^\star) = \min_{P\in K_{\Tcal,\widehat X}}\widetilde F(P).
\]
Additionally, if the optimizer of \eqref{eq:discretized_WOT} is unique, then $P^\star=P^*$.
\end{theorem}

\begin{proof}
We work on $\aff(K_{\Tcal,\widehat X})$, and $\partial\widetilde F$ denotes the corresponding relative subdifferential. As components orthogonal to its direction space do not affect the projected updates, we may restrict attention to relative subgradients. By assumption, $\widetilde F$ is finite on $K_{\Tcal,\widehat X}$.

Since $\widetilde F$ is convex, it is therefore locally Lipschitz on $\ri(K_{\Tcal,\widehat X})$. As $\overline{K_{\Tcal,\widehat X}^\varepsilon} \subseteq \ri(K_{\Tcal,\widehat X})$ is compact, there exists $M_\varepsilon>0$ such that $\|G\|\leq M_\varepsilon$ for all $G\in\partial\widetilde F(P)$ and $P\in K_{\Tcal,\widehat X}^\varepsilon$. In particular, $\|G^{(k)}\|\leq M_\varepsilon$ for every $k\in\NN$.

By the non-expansiveness of the Euclidean projection and the subgradient inequality,
\begin{align}
\|P^{(k+1)} - P^*\|^2_{HS}
& \leq \|P^{(k)} - \tau_kG^{(k)} - P^*\|^2_{HS} \nonumber\\
& \leq \|P^{(k)} - P^*\|^2_{HS} - 2\tau_k\big(\widetilde F(P^{(k)}) - \widetilde F(P^*)\big) + M_\varepsilon^2\tau_k^2.
\label{eq:projected-subgradient-estimate}
\end{align}
Summing \eqref{eq:projected-subgradient-estimate} from $k=0$ to $N$ gives
\[
2\sum_{k=0}^N\tau_k\big(\widetilde F(P^{(k)}) - \widetilde F(P^*)\big)
\leq \|P^{(0)} - P^*\|^2_{HS} + M_\varepsilon^2\sum_{k=0}^N\tau_k^2.
\]
Since $P^*$ is an optimizer, $\widetilde F(P^{(k)}) - \widetilde F(P^*) \geq 0$, for all $k\in\NN$. Therefore, $\sum_{k=0}^\infty\tau_k^2<\infty$ yields $\sum_{k=0}^\infty\tau_k\big(\widetilde F(P^{(k)}) - \widetilde F(P^*)\big) < \infty$. Since $\sum_{k=0}^\infty\tau_k=\infty$, it follows that
\[
\liminf_{k\to\infty}\big(\widetilde F(P^{(k)}) - \widetilde F(P^*)\big) = 0.
\]
Indeed, otherwise there would exist $\delta>0$ and $k_0\in\NN$ such that $\widetilde F(P^{(k)}) - \widetilde F(P^*) \geq \delta$ for $k\geq k_0$, which would imply $\sum_{k=k_0}^\infty\tau_k\big(\widetilde F(P^{(k)}) - \widetilde F(P^*)\big) \geq \delta\sum_{k=k_0}^\infty\tau_k = \infty$, a contradiction. Therefore,
\[
\liminf_{k\to\infty}\widetilde F(P^{(k)}) = \widetilde F(P^*).
\]
Hence, by compactness of $K_{\Tcal,\widehat X}^\varepsilon$, there exists a subsequence $P^{(k_\ell)} \to P^\star \in K_{\Tcal,\widehat X}^\varepsilon$ such that, by continuity of $\widetilde F$, $\widetilde F(P^\star) = \widetilde F(P^*)$.

Applying \eqref{eq:projected-subgradient-estimate} with $P^\star$ in place of $P^*$ yields $\|P^{(k+1)} - P^\star\|^2_{HS} \leq \|P^{(k)} - P^\star\|^2_{HS} + M_\varepsilon^2\tau_k^2$. Therefore, $\|P^{(k)} - P^\star\|^2_{HS} + M_\varepsilon^2\sum_{j=k}^\infty\tau_j^2$ is nonincreasing, and hence $(\|P^{(k)} - P^\star\|_{HS})_{k\in\NN}$ admits a limit. Since $P^{(k_\ell)}\to P^\star$, this limit is zero, and therefore $P^{(k)}\to P^\star$. The convergence of the objective values follows by continuity of $\widetilde F$. If the optimizer is unique, necessarily $P^\star=P^*$.
\end{proof}

\begin{remark}[Further parallelization]
An additional level of parallelization may be obtained by exploiting Proposition~\ref{prop:decomposition-fixed-marginals}. Indeed, keeping the odd-indexed marginals fixed, the optimization over the even-indexed marginals decomposes into independent subproblems that can be solved in parallel. The same applies to the odd-indexed marginals when the even-indexed ones are fixed. This suggests an alternating optimization procedure, which can be implemented using different numerical methods. Although convergence to a global optimizer is not ensured in general, this additional level of parallelization may offer computational advantages when the number of elements of the partition $\Tcal$ is large.
\end{remark}

\section{Further extensions and applications}\label{sec:extensions}

\paragraph{Calibration with relaxed constraints.}The calibration problems considered so far require the prescribed market constraints to be satisfied exactly. In practice, however, observed derivative prices may be affected by noise or be mutually inconsistent, potentially leaving the admissible set $\Mart_\Tcal$ empty. To overcome this issue, we can introduce a relaxed optimization problem in which the calibration constraints are incorporated into the objective function through a penalty term.

Let $\Mcal_\Tcal^{\mathrm{all}} := \left\{(\eta_i)_{i=1}^{n+1} \in \Prob_2(A)^{n+1}:\eta_i \preceq_c \eta_{i+1}, \ i = 1, \dots, n\right\}$. Suppose that the constraints on a chain $\eta = (\eta_i)_{i=1}^{n+1} \in \Mcal_\Tcal^{\mathrm{all}}$ can be written as $f(\eta)=0$ for some map $f:\Mcal_\Tcal^{\mathrm{all}} \to \RR^N$. Rather than restricting the optimization to chains satisfying these constraints exactly, we consider
\begin{equation}
\label{eq:relaxed_WOT}
\inf_{\eta\in\Mcal_\Tcal^{\mathrm{all}}}\left\{F(\eta) + \lambda\,\Psi(f(\eta))\right\},
\end{equation}
where $\Psi:\RR^N\to[0,\infty)$ penalizes violations of the prescribed constraints, and $\lambda>0$ controls the strength of the penalization.

In particular, if $\Psi\circ f$ is convex and finite on $\Mcal_\Tcal^{\mathrm{all}}$, then \eqref{eq:relaxed_WOT} remains a convex optimization problem. Therefore, we can apply the same discretization introduced in Section~\ref{sec:numerical-algorithm}, with $K_{\Tcal,\widehat X}$ replaced by the corresponding feasible set $K_{\Tcal,\widehat X}^{\mathrm{all}}$ associated with $\Mcal_\Tcal^{\mathrm{all}}$. It is then enough to replace the subgradient $G^{(k)}$ by $G^{(k)}+H^{(k)}$, where
\[
H^{(k)} \in \partial(\lambda\Psi\circ f\circ\eta)(P^{(k)}).
\]

\begin{example}[Calibration to call option prices]
Consider the calibration to $N$ call option prices. For each $h=1,\dots,N$, let $K_h$, $c_h$ and $s_{i_h}\in\Tcal$ denote the strike, observed price and maturity of the $h$-th option, respectively. Define
\[
f_h(P) := \sum_{j=1}^m(\widehat x_{i_hj} - K_h)^+p_{i_hj} - c_h, \qquad f(P) := (f_h(P))_{h=1}^N.
\]
Choosing $\Psi(z)=|z|$, the Euclidean norm on $\RR^N$, the relaxed objective becomes $\widetilde F(P)+\lambda|f(P)|$.

If the observed prices are inconsistent with the martingale constraints, then $f(P)\neq 0$ for every $P \in K_{\Tcal,\widehat X}^{\text{all}}$. In this case, a subgradient of the penalty term at $P^{(k)}$ is given by the matrix $H^{(k)}=(H_{ij}^{(k)})$, where
\[
H_{ij}^{(k)} = \frac{\lambda}{|f(P^{(k)})|}\sum_{\{h:i_h=i\}}f_h(P^{(k)})(\widehat x_{ij} - K_h)^+, \qquad \text{if $|f(P^{(k)})| \neq 0$},
\]
$H_{ij}^{(k)}=0$, otherwise. 
\end{example}

\paragraph{Maximizing the maximum covariance.} The approach developed in this paper for \eqref{eq:mainPb} can also be adapted to the related dynamic problem
\begin{equation}
\label{eq:dynamic_MOT2}
\widetilde{MOT}(\Mart_{[t_0,t_1]}, \widehat\sigma) := \sup_{M\in\Mart_{[t_0,t_1]}} \EE\left[\int_{t_0}^{t_1}\langle\sigma_t, \widehat\sigma(t, M_t)\rangle_{HS}\,dt\right].
\end{equation}
More precisely, one can construct an analogous discrete approximation by considering
\begin{equation}
\label{eq:WOT2}
\widetilde{MWT}\big(\Mcal_\Tcal, (q_\cdot^{(i)})_{i=1}^n\big) := \sup_{(\eta_i)_{i=1}^{n+1}\in\Mcal_\Tcal}\sum_{i=1}^n \left(WT_{\Scal}^{q_\cdot^{(i)}}(\eta_i, \eta_{i+1}) - \int_\RRd |y|^2 \eta_{i}(dy)\right),
\end{equation}
where
\[
WT_{\Scal}^{q_\cdot}(\rho, \zeta) := \sup_{\pi\in\Mart(\rho,\zeta)}\int_{\RRd}\MCov(\pi_x, q_x)\,\rho(dx).
\]

To motivate this approximation, let $M^{(i)}$ be a continuous-time $q_x^{(i)}$-Bass extension and $\widehat M^{(i)}$ its associated reference martingale. Since both processes start from the same point, Itô's isometry yields
\[
WT_{\Scal}^{q_\cdot^{(i)}}(\eta_i, \eta_{i+1}) - \int_\RRd |y|^2\eta_i(dy) = \EE\left[\int_{s_i}^{s_{i+1}}\langle\sigma_r^{(i)}, \widehat\sigma(r, \widehat M_r^{(i)})\rangle_{HS}\,dr\right].
\]
As the mesh tends to zero, replacing $\widehat M_r^{(i)}$ by $M_r^{(i)}$ in the reference coefficient motivates the approximation of \eqref{eq:dynamic_MOT2} by \eqref{eq:WOT2}.

An analogue of Proposition~\ref{prop:approximate-problem} can then be established for \eqref{eq:WOT2}, with concavity replacing convexity, provided the supremum is finite. In particular, finiteness is ensured if the admissible marginal chains are uniformly compactly supported and the reference kernels have uniformly bounded first moments on their supports.

For fixed marginals $\rho$ and $\zeta$, the problems $WT_{\Scal}^{q_\cdot}(\rho,\zeta)$ and $WT_{\Ical}^{q_\cdot}(\rho,\zeta)$ have the same optimal martingale coupling; see \cite[Proposition~2.6]{Ma26_qxBass}. However, when optimizing over the entire chain of marginals, \eqref{eq:WOT} and~\eqref{eq:WOT2} may select different optimal chains. Nonetheless, the duality theory for $q_x$-Bass martingales can be used in the same way to construct a maximization algorithm for \eqref{eq:WOT2}.

\paragraph{Applications to the Martingale Benamou--Brenier problem and its generalizations.}
When the initial and terminal marginals are fully prescribed and the reference is an additive martingale with continuous sample paths, the present framework is not needed. Indeed, the independence of its increments ensures that the conditional-expectation interpolation associated with the Bass construction is also optimal for the intermediate marginals it generates. This is the case, for instance, of the Martingale Benamou--Brenier problem \cite{BaBeHuKa20}, where the reference is Brownian motion, and its generalization to $X$-Bass martingales, where the reference process $X$ has independent increments \cite{AcMa26_general}.

However, when only partial marginal information is available, for instance through finitely many vanilla option prices, our approach can be applied to select an optimal chain of marginal distributions so that the associated Bass interpolation has dynamics as close as possible to those of the prescribed reference process, among all martingales satisfying the constraints. More specifically, in this case, it suffices to consider a partition containing only the maturities at which constraints are imposed and to solve \eqref{eq:WOT} for the optimal chain of marginals. The corresponding continuous-time optimizer for \eqref{eq:mainPb} can then be constructed by joining the appropriate Bass interpolations between the marginals of the optimal chain, according to the chosen reference process.

It is important to remark that, in this setting, where the complete marginal distributions are not prescribed, the optimization problem \eqref{eq:mainPb} is not equivalent to the maximal covariance formulation \eqref{eq:dynamic_MOT2}, which can however be solved similarly by considering the discrete-time problem \eqref{eq:WOT2}.

We refer to \cite{KoNoSo26} for a related theoretical study of the dual problem of the martingale Benamou--Brenier formula in a one-dimensional setting with call-price constraints under suitable assumptions.

\paragraph{When the reference includes additional factors.}
Although the natural application of the multidimensional setting is in principle the joint calibration of several asset prices, provided sufficiently rich market data are available, additional dimensions can also be used to introduce latent sources of randomness while calibrating only the asset-price component. 

Inspired by the stochastic-volatility framework considered in \cite{GuLoWa22}, consider, under the risk-neutral measure and with discounted asset price, the two-factor model
\[
dS_t = S_t\varsigma(t, \log S_t, V_t)\,dW_t^1,
\qquad
dV_t = \kappa(\theta - V_t)\,dt + \xi\sqrt{V_t}\,dW_t^2,
\]
where $d\langle W^1, W^2\rangle_t = \eta(t, \log S_t, V_t)\,dt$. Here, $V$ represents a latent stochastic-volatility factor, which need not coincide with the instantaneous relative variance $\varsigma^2(t,\log S_t,V_t)$ of the asset price. Although the joint process $(S,V)$ is Markovian, the factor $V$ is not a martingale, as required by our framework. This can be addressed by introducing $Y_t:=e^{\kappa t}(V_t-\theta)$, which satisfies, by Itô's formula,
\[
dY_t = \xi e^{\kappa t}\sqrt{V_t}\,dW_t^2.
\]
Hence, $\widehat M=(S,Y)$ is a two-dimensional Markov martingale. Since $V_t=\theta+e^{-\kappa t}Y_t$, the transformation preserves the information contained in the original state.

Under suitable regularity and non-degeneracy conditions, the corresponding transition kernels are absolutely continuous and the approach based on the continuous-time $(q_x^{(i)})_{i=1}^n$-Bass interpolation of the optimizers of \eqref{eq:WOT} can be applied with this reference, imposing calibration constraints only on $S$ and leaving $Y$ unconstrained by market prices. Unlike \cite{GuLoWa22}, where the dynamics of the stochastic-volatility factor and the covariance structure are prescribed in the calibration problem, this optimization does not impose such structural restrictions on the additional component. Instead, its dynamics are selected jointly with those of the asset price. Consequently, although the additional component represents a transformed volatility factor in the reference, its dynamics need not be preserved by the optimizer, and it may be interpreted as a latent stochastic factor in the calibrated model.

\appendix
\section{Appendix}

\begin{lemma}
\label{lem:one-step-estimate}
Let $t_0\leq s<t\leq t_1$ and set $h:=t-s$. Suppose that \ref{ass:A1} and~\ref{ass:A2} hold. Let $N$ be an $A$-valued Brownian martingale on $[s,t]$ of the form
\[
	N_r = N_s + \int_s^r\sigma^N_u\,dB_u, \qquad r \in [s, t],
\]
where $\sigma^N$ is predictable and $\rho := \Law(N_s), \zeta := \Law(N_t) \in \Prob_2(A)$. Then there exists $C>0$, depending only on $L$ and $t_1-t_0$, such that
\[
	WT_{\Ical}^{q_\cdot^{s,t}}(\rho, \zeta) \leq (1 + C\sqrt h)\EE\left[\int_s^t\|\sigma^N_r - \widehat\sigma(r, N_r)\|_{HS}^2\,dr\right].
\]
If, in addition, $N$ is the continuous-time $q_x^{s,t}$-Bass extension associated with a solution $\widehat N$ of \eqref{eq:reference_diffusion}, then
\[
	\EE\left[\int_s^t\|\sigma^N_r - \widehat\sigma(r, N_r)\|_{HS}^2\,dr\right] \leq (1 + C\sqrt h) WT_{\Ical}^{q_\cdot^{s,t}}(\rho, \zeta).
\]
\end{lemma}

\begin{proof}
Let $\widehat N$ be the strong solution of $d\widehat N_r = \widehat\sigma(r, \widehat N_r)\,dB_r$, with initial condition $\widehat N_s=N_s$, driven by the same Brownian motion as $N$. Writing $\Law(N_s, N_t)(dx, dy) = \rho(dx)\pi_x(dy)$, we have $\pi\in\Mart(\rho,\zeta)$, while $\Law(\widehat N_t|N_s=x)= q_x^{s,t}$. Hence,
\begin{equation}
\label{eq:WT_estimate}
	WT_{\Ical}^{q_\cdot^{s,t}}(\rho, \zeta) \leq \EE[|N_t - \widehat N_t|^2] = \EE\left[\int_s^t\|\sigma^N_r - \widehat\sigma(r, \widehat N_r)\|_{HS}^2\,dr\right],
\end{equation}
where the last equality follows from Itô's isometry. Moreover,
\begin{equation}
\label{eq:decomposition_diffusion}
	\sigma^N_r - \widehat\sigma(r, \widehat N_r) = (\sigma^N_r - \widehat\sigma(r, N_r)) + (\widehat\sigma(r, N_r) - \widehat\sigma(r, \widehat N_r)).
\end{equation}
Then \eqref{eq:WT_estimate},  \eqref{eq:decomposition_diffusion} and~\ref{ass:A2} imply that, for every $r\in[s,t]$,
\[
	\EE[|N_r - \widehat N_r|^2] \leq 2\EE\left[\int_s^r\|\sigma^N_u - \widehat\sigma(u, N_u)\|_{HS}^2\,du\right] + 2L^2\int_s^r\EE[|N_u - \widehat N_u|^2]\,du.
\]
Therefore, Gronwall's inequality yields
\[
	\sup_{r\in[s,t]}\EE[|N_r - \widehat N_r|^2] \leq C'\EE\left[\int_s^t\|\sigma^N_r - \widehat\sigma(r, N_r)\|_{HS}^2\,dr\right],
\]
where $C':=2\exp\big(2L^2(t_1-t_0)\big)$. Using again \ref{ass:A2} and the previous estimate, we obtain
\begin{equation}
\label{eq:first_estimate}
\begin{aligned}
\EE\left[\int_s^t\|\widehat\sigma(r, N_r) - \widehat\sigma(r, \widehat N_r)\|_{HS}^2\,dr\right]
& \leq L^2\int_s^t\EE[|N_r - \widehat N_r|^2]\,dr \\
& \leq L^2C'h\,\EE\left[\int_s^t\|\sigma^N_r - \widehat\sigma(r, N_r)\|_{HS}^2\,dr\right].
\end{aligned}
\end{equation}

Therefore, \eqref{eq:decomposition_diffusion} and \eqref{eq:first_estimate} yield
\[
	\EE\left[\int_s^t\|\sigma^N_r - \widehat\sigma(r, \widehat N_r)\|_{HS}^2\,dr\right] \leq \left(1 + L\sqrt{C'h}\right)^2\EE\left[\int_s^t\|\sigma^N_r - \widehat\sigma(r, N_r)\|_{HS}^2\,dr\right],
\]
since, by the Cauchy--Schwarz inequality and \eqref{eq:first_estimate}, we have
\begin{align*}
	2 & \left|\EE\left[\int_s^t\left\langle \sigma^N_r - \widehat\sigma(r, N_r), \widehat\sigma(r, N_r) - \widehat\sigma(r, \widehat N_r)\right\rangle_{HS}\,dr\right]\right|\\
	& \qquad\qquad\qquad\qquad\qquad\qquad\qquad\qquad \leq 2L\sqrt{C'h}\,\EE\left[\int_s^t\|\sigma^N_r - \widehat\sigma(r, N_r)\|_{HS}^2\,dr\right].
\end{align*}

As $h\leq t_1-t_0$, we have $h\leq\sqrt{t_1-t_0}\sqrt h$, and therefore $\left(1 + L\sqrt{C'h}\right)^2 \leq 1 + C\sqrt h$, where $C:=2L\sqrt{C'}+L^2C'\sqrt{t_1-t_0}$. Hence,
\[
	\EE\left[\int_s^t\|\sigma^N_r - \widehat\sigma(r, \widehat N_r)\|_{HS}^2\,dr\right] \leq (1 + C\sqrt h)\EE\left[\int_s^t\|\sigma^N_r - \widehat\sigma(r, N_r)\|_{HS}^2\,dr\right].
\]
Combining this estimate with \eqref{eq:WT_estimate} proves the first claim.

Assume now that $N$ is the continuous-time $q_x^{s,t}$-Bass extension associated with $\widehat N$. Then, the coupling of the conditional laws of $N_t$ and $\widehat N_t$ induced by their joint law is optimal, and therefore
\begin{equation}
\label{eq:WOT_value_representation}
	WT_{\Ical}^{q_\cdot^{s,t}}(\rho, \zeta) = \EE[|N_t - \widehat N_t|^2] = \EE\left[\int_s^t\|\sigma^N_r - \widehat\sigma(r, \widehat N_r)\|_{HS}^2\,dr\right].
\end{equation}
Moreover, since $N-\widehat N$ is a martingale, the process $|N-\widehat N|^2$ is a submartingale, and thus, for every $r\in[s,t]$,
\begin{equation}
\label{eq:WOT_value_inequality1}
\EE[|N_r - \widehat N_r|^2] \leq \EE[|N_t - \widehat N_t|^2] = WT_{\Ical}^{q_\cdot^{s,t}}(\rho, \zeta).
\end{equation}
Hence, by \ref{ass:A2},
\begin{equation}
\label{eq:WOT_value_inequality2}
\EE\left[\int_s^t\|\widehat\sigma(r, N_r) - \widehat\sigma(r, \widehat N_r)\|_{HS}^2\,dr\right] \leq L^2\int_s^t\EE[|N_r - \widehat N_r|^2]\,dr \leq L^2h WT_{\Ical}^{q_\cdot^{s,t}}(\rho, \zeta).
\end{equation}

Now, consider the decomposition $\sigma^N_r - \widehat\sigma(r, N_r) = (\sigma^N_r - \widehat\sigma(r, \widehat N_r)) + (\widehat\sigma(r, \widehat N_r) - \widehat\sigma(r, N_r))$. By the Cauchy--Schwarz inequality and \eqref{eq:WOT_value_representation}--\eqref{eq:WOT_value_inequality2},
\[
	2\left|\EE\left[\int_s^t\left\langle\sigma^N_r - \widehat\sigma(r, \widehat N_r), \widehat\sigma(r, \widehat N_r) - \widehat\sigma(r, N_r)\right\rangle_{HS}\,dr\right]\right| \leq 2L\sqrt h WT_{\Ical}^{q_\cdot^{s,t}}(\rho, \zeta).
\]
Therefore, expanding the square and using the previous bound, we obtain
\[
	\EE\left[\int_s^t\|\sigma^N_r - \widehat\sigma(r, N_r)\|_{HS}^2\,dr\right] \leq (1 + L\sqrt h)^2 WT_{\Ical}^{q_\cdot^{s,t}}(\rho, \zeta).
\]
Since $(1+L\sqrt h)^2\leq1+C\sqrt h$ for the constant $C$ defined above, the second claim follows.
\end{proof}

\begin{proof}[Proof of Proposition~\ref{prop:finite-support-approximation}]
Set
\[
I := \inf_{\substack{m\in\NN\\ \widehat X\in\mathfrak X_m}}\ \inf_{\eta\in\Mcal_{\Tcal,\widehat X}}F(\eta).
\]
Since $\Mcal_{\Tcal,\widehat X} \subseteq \Mcal_\Tcal$ for every $m\in\NN$ and $\widehat X\in\mathfrak X_m$, we immediately have $MWT\big(\Mcal_\Tcal, (q_\cdot^{(i)})_{i=1}^n\big) \leq I$. For the reverse inequality, fix $\eta = (\eta_i)_{i=1}^{n+1} \in \Mcal_\Tcal$. For every $i=1,\dots,n$, let $\pi^{(i)}$ be the optimizer of $WT_{\Ical}^{q_\cdot^{(i)}}(\eta_i, \eta_{i+1})$. By Proposition~\ref{prop:approximate-problem}, their Markovian gluing defines a discrete-time martingale $(X_i)_{i=1}^{n+1}$ such that $\Law(X_i)=\eta_i$ and $\Law(X_i,X_{i+1})=\pi^{(i)}$.

For every $k\in\NN$ and $i=1,\dots,n+1$, let $\Gcal_i^{(k)}$ be a finite sub-$\sigma$-algebra of $\sigma(X_1,\dots,X_i)$ such that $\Gcal_i^{(k)} \subseteq \Gcal_{i+1}^{(k)}$, $\Gcal_i^{(k)} \subseteq \Gcal_i^{(k+1)}$, and $\bigvee_{k\in\NN}\Gcal_i^{(k)} = \sigma(X_1, \dots, X_i)$. 
Furthermore, recall that, for each $s_i\in\Tcal$, the constraints on $X_i$ can be expressed as equalities or inequalities involving expectations of continuous payoff functions that are affine on finitely many convex regions covering $A$. We therefore require that $\{X_i\in C\}\in\Gcal_i^{(k)}$ for every such region $C$ and every $k\in\NN$.

Now, set
\[
X_i^{(k)} := \EE[X_i\mid\Gcal_i^{(k)}], \qquad \eta_i^{(k)} := \Law(X_i^{(k)}), \qquad i = 1, \dots, n + 1.
\]
Since $A$ is convex and $X_i\in A$ a.s., we have $X_i^{(k)}\in A$ a.s. Moreover,
\[
\EE[X_{i+1}^{(k)}\mid\Gcal_i^{(k)}] = \EE[X_{i+1}\mid\Gcal_i^{(k)}] = \EE[X_i\mid\Gcal_i^{(k)}] = X_i^{(k)},
\]
so $(X_i^{(k)})_{i=1}^{n+1}$ is a martingale and therefore $\eta_i^{(k)} \preceq_c \eta_{i+1}^{(k)}$ for every $i=1,\dots,n$. 
Additionally, the prescribed marginal constraints are preserved by construction, since the chosen sub-$\sigma$-algebras contain the events $\{X_i\in C\}$ corresponding to the convex regions $C$ on which the prescribed payoff functions are affine.

Thus $\eta^{(k)} := (\eta_i^{(k)})_{i=1}^{n+1} \in \Mcal_\Tcal$. Since each $\Gcal_i^{(k)}$ is finite, each $\eta_i^{(k)}$ is supported on finitely many atoms. Taking the union of these supports, we obtain a common finite grid in $A$. Since $A$ is open and convex, by adding finitely many zero-mass atoms at each time step, we can construct $\widehat X^{(k)}\in\mathfrak X_{m_k}$ satisfying \eqref{eq:support_irreducible}, without changing the marginals. Hence $\eta^{(k)} \in \Mcal_{\Tcal,\widehat X^{(k)}}$.

Since $X_i\in L^2$ and $\Gcal_i^{(k)}\uparrow\sigma(X_1, \dots, X_i)$, the martingale convergence theorem yields $X_i^{(k)}\to X_i$ in $L^2$, and consequently $\Wcal_2(\eta_i^{(k)},\eta_i)\to 0$ as $k\to\infty$, for any $i\in\{1,\dots,n+1\}$.

Fix $i\in\{1,\dots,n\}$ and write $\pi^{(i)}(dx, dy) = \eta_i(dx)\pi_x^{(i)}(dy)$. On an enlargement of the probability space, choose $Z_{i+1}$ such that, conditionally on $X_i=x$, $(X_{i+1},Z_{i+1})$ is an optimal coupling of $\pi_x^{(i)}$ and $q_x^{(i)}$. Then
\[
	\EE[|X_{i+1} - Z_{i+1}|^2] = WT_{\Ical}^{q_\cdot^{(i)}}(\eta_i, \eta_{i+1}).
\]
By Proposition \ref{prop:reference-transition-kernel}, the map $x \mapsto q_x^{(i)}$ is Lipschitz with respect to the $\Wcal_2$ distance. Therefore, there exists $K_i>0$ such that $\Wcal_2(q_x^{(i)}, q_y^{(i)}) \leq K_i|x - y|$ for all $x,y\in A$. Taking a measurable family of optimal couplings between $q_x^{(i)}$ and $q_{x'}^{(i)}$, with a conditional gluing, we may construct $Z_{i+1}^{(k)}$ such that $\Law(Z_{i+1}^{(k)}\mid X_i^{(k)}) = q_{X_i^{(k)}}^{(i)}$ and $\EE[|Z_{i+1}^{(k)} - Z_{i+1}|^2] \leq K_i^2\EE[|X_i^{(k)} - X_i|^2].$
 
Since $\Law(X_i^{(k)},X_{i+1}^{(k)})$ is a martingale coupling of $\eta_i^{(k)}$ and $\eta_{i+1}^{(k)}$, the conditional law of $(X_{i+1}^{(k)},Z_{i+1}^{(k)})$ given $X_i^{(k)}$ is admissible in the corresponding one-step problem. Hence
\[
	\sqrt{WT_{\Ical}^{q_\cdot^{(i)}}(\eta_i^{(k)}, \eta_{i+1}^{(k)})} \leq \|X_{i+1}^{(k)} - Z_{i+1}^{(k)}\|_{L^2} \leq \|X_{i+1}^{(k)} - X_{i+1}\|_{L^2} + \sqrt{WT_{\Ical}^{q_\cdot^{(i)}}(\eta_i, \eta_{i+1})} + K_i\|X_i^{(k)} - X_i\|_{L^2}.
\]
Letting $k\to\infty$ gives $\limsup_{k\to\infty}WT_{\Ical}^{q_\cdot^{(i)}}(\eta_i^{(k)}, \eta_{i+1}^{(k)}) \leq WT_{\Ical}^{q_\cdot^{(i)}}(\eta_i, \eta_{i+1})$.

Summing over $i=1,\dots,n$ yields $\limsup_{k\to\infty}F(\eta^{(k)}) \leq F(\eta)$. Since $\eta^{(k)} \in \Mcal_{\Tcal,\widehat X^{(k)}}$, we have $I\leq F(\eta^{(k)})$ for every $k$, and therefore $I\leq F(\eta)$. Taking the infimum over $\eta\in\Mcal_\Tcal$ gives
\[
I \leq MWT\big(\Mcal_\Tcal, (q_\cdot^{(i)})_{i=1}^n\big).
\]
\end{proof}

\printbibliography
\end{document}